\documentclass[lettersize,journal]{IEEEtran}
\usepackage{amsmath,amsfonts}
\usepackage{array}
\usepackage[caption=false,font=normalsize,labelfont=sf,textfont=sf]{subfig}
\usepackage{textcomp}
\usepackage{stfloats}
\usepackage{url}
\usepackage{verbatim}
\usepackage{graphicx}
\usepackage{cite}
\usepackage{cuted}
\usepackage{amsmath,amssymb,amsfonts}
\usepackage{graphicx}
\usepackage{textcomp}
\usepackage{soul}
\usepackage{setspace}
\usepackage{amsthm}
\usepackage{todonotes}
\usepackage{menukeys}
\usepackage{soul}
\usepackage{bbm}

\usepackage[compact]{titlesec}
\titlespacing{\section}{0pt}{2ex}{1ex}
\titlespacing{\subsection}{0pt}{2ex}{1ex}
\usepackage[table]{xcolor}
\definecolor{darkgreen}{rgb}{0,0.6,0}
\usepackage{tabularx}
\usepackage{nccmath}
\usepackage[strict]{changepage}

\usepackage{cite}
\usepackage[colorlinks=true, allcolors=blue]{hyperref}
\usepackage{tikz}
\usepackage{setspace}
\usetikzlibrary{arrows,shapes}
\usepackage{nicematrix}
\usepackage{arydshln}

\usepackage{float}           
\usepackage{algorithm}
\usepackage{algorithmicx}
\usepackage[noend]{algpseudocode}
\usepackage{arydshln}
\definecolor{green}{RGB}{0,0,0}

\renewcommand{\qedsymbol}{$\blacksquare$}

\makeatletter
\newcommand{\manuallabel}[2]{\phantomsection\def\@currentlabel{#2}\label{#1}}
\makeatother

\usepackage{titlesec}
\titlespacing*{\subsection}{0pt}{0.2\baselineskip}{0.2\baselineskip} 

\definecolor{der1}{RGB}{0,114,189}   
\definecolor{der2}{RGB}{217,83,25}   
\definecolor{der3}{RGB}{237,177,32}  
\definecolor{der4}{RGB}{126,47,142}  
\definecolor{der5}{RGB}{119,172,48}  

\newcommand{\legline}[2]{%
  \textcolor{#1}{\raisebox{0.4ex}{\rule{1.6em}{2.5pt}}}\,#2%
}

\usepackage{etoolbox}

\makeatletter
\expandafter\patchcmd\csname\string\IEEEbiography\endcsname
  {\vskip \@IEEEBIOskipN plus 1fil minus 0\baselineskip}
  {\vskip \@IEEEBIOskipN}
  {}{}
\makeatother

\begin{document}

\title{A Connectively Stable and Robust \textcolor{black}{Distributed Frequency and Voltage} Control Scheme for Islanded Networks of Microgrids}

\author{Ahmed Saad Al-Karsani,~\IEEEmembership{Student Member,~IEEE,} Maryam Khanbaghi,~\IEEEmembership{Senior Member,~IEEE}, Aleksandar Zečević,~\IEEEmembership{Senior Member,~IEEE}
\thanks{Ahmed Saad Al-Karsani, Maryam Khanbaghi (corresponding author) and Aleksandar Zečević are with Santa Clara University, 500 El Camino Real, Santa Clara, CA, USA. (email: \{ahsaad, mkhanbaghi, azecevic\}@scu.edu)}
}

\markboth{ }%
{Shell \MakeLowercase{\textit{et al.}}: A Sample Article Using IEEEtran.cls for IEEE Journals}


\maketitle

\begin{abstract}
The transition towards clean energy and the introduction of Distributed Energy Resources (DERs) are giving rise to the emergence of Microgrids (MGs) and Networks of MGs (NMGs). MGs and NMGs can operate autonomously in islanded mode. However, they face challenges in terms of secondary level frequency and voltage regulation, due to the variable nature of Renewable Energy Sources (RES) and loads. Distributed-Averaging Proportional-Integral (DAPI) control has been proposed in the literature for distributed frequency and voltage control of droop-controlled DERs, but it {\color{black}might not guarantee robustness to cyber-physical disturbances}. To address this, we propose a robust DAPI frequency and voltage control scheme that ensures robustness using the concept of connective stability, along with the invariant ellipsoid technique for disturbance rejection. {\color{black}The proposed scheme adds additional degrees of freedom to determine the strength of distributed control, providing an option for more robust action}. Simulation of an NMG model in MATLAB\textsuperscript{\textregistered}/Simulink\textsuperscript{\textregistered} consisting of 3 MGs and 5 DERs {\color{black}shows that the proposed scheme minimizes frequency and voltage deviations during cyber-physical disturbance events compared to the baseline, dissipativity-based and $H_\infty$-based techniques.}
\end{abstract}


{\color{black}\section*{Nomenclature}

\newcommand{\NomWidest}{\textbf{$A,B,C,D$}}

\subsection*{Abbreviations}
\begin{IEEEdescription}[\IEEEsetlabelwidth{$\Delta \mathbf{Q},\Delta \mathbf{R},\Delta $}\IEEEusemathlabelsep]
  \item[ACE] Area Control Error
  \item[AGC] Automatic Generation Control
  \item[BESS] Battery Energy Storage System
  \item[DAPI] Distributed-Averaging Proportional-Integral
  \item[DER] Distributed Energy Resource
  \item[DoS] Denial of Service
  \item[EIP] Equilibrium-Independent Passivity
  \item[FDI] False Data Injection
  \item[GFL] {\color{green}Grid Following}
  \item[GFM] Grid Forming
  \item[IBR] Inverter-Based Resource
  \item[ISS] Input-to-State Stability
  \item[LMI] Linear Matrix Inequality
  \item[LPF] Low-Pass Filter
  \item[MAS] Multi-Agent Systems
  \item[MG] Microgrid
  \item[NMG] Network of Microgrids
  \item[PV] {\color{green}Photovoltaic}
  \item[RES] Renewable Energy Sources
  \item[PCC] Point of Common Coupling
  \item[PWM] Pulse Width Modulation
\end{IEEEdescription}

\subsection*{Symbols}
\begin{IEEEdescription}[\IEEEsetlabelwidth{$\Delta \mathbf{Q},\Delta \mathbf{R},\Delta $}\IEEEusemathlabelsep]
  \item[$\bar{(\cdot)}$] Bound
  \item[$(\cdot)^*$] Setpoint value
  \item[$(\cdot)_D$] NMG composite expression
  \item[$(\cdot)_{i,j}$] $i^{th},j^{th}$ DER expression
\end{IEEEdescription}

\subsection*{Parameters}
\begin{IEEEdescription}[\IEEEsetlabelwidth{$\Delta \mathbf{Q},\Delta \mathbf{R},\Delta$}\IEEEusemathlabelsep]
  \item[$A,B,E$] State, input, disturbance matrix
  \item[$\Delta A, \Delta E$] State, disturbance uncertainty matrix
  \item[$a,b$] Frequency, voltage consensus communication link gain
  \item[$\bar{\boldsymbol{e}}_{ij}$] Interconnection fundamental strength
  \item[$G,H$] Disturbance, state uncertainty matrix
  \item[$k_{i},\kappa_{i}$] Frequency, voltage consensus inverse integral gain
  \item[$m,n$] Frequency (rad/Ws), voltage (V/VAR) droop gain
  \item[$M$] Total number of NMG inputs 
  \item[$N$] Total number of NMG DERs 
  \item[$\mathbf{Q},\mathbf{R},\mathbf{S}$] Disturbance, disturbance-state, state weighting supply rate matrix
  \item[$\Delta \mathbf{Q},\Delta \mathbf{R},\Delta \mathbf{S}$] Disturbance, disturbance-state, state weighting supply rate uncertainty matrix
  \item[$K,L$] State-feedback, transformed gain matrix 
  \item[$\boldsymbol{P},Y$] Lyapunov, inverse Lyapunov matrix
  \item[$\alpha,\beta$] Frequency, voltage consensus connective strength
  \item[$\gamma_\alpha,\gamma_\beta$] Frequency, voltage consensus inverse square connective strength
  \item[$\kappa_L$] State feedback bound
  \item[$\xi$] Consensus voltage deviation term gain
  \item[$\tau$] S-procedure multiplier
  \item[$\tau_c$] Low-pass filter time constant (sec)
\end{IEEEdescription}

\subsection*{Variables}
\begin{IEEEdescription}[\IEEEsetlabelwidth{$\Delta \mathbf{Q},\Delta \mathbf{R},\Delta$}\IEEEusemathlabelsep]
  \item[$\boldsymbol{E}$] Interconnection matrix
  \item[$\boldsymbol{e}_{ij}$] Interconnection actual strength
  \item[$e$] Voltage consensus (V)
  \item[$g(d),h(x)$] Disturbance, state uncertainty function
  \item[$P,Q$] Average active (W), reactive (VAR) power
  \item[$p,q$] Instantaneous active (W), reactive (VAR) power 
  \item[$t$] Time (sec)
  \item[$V$] Voltage (V)
  \item[$x,u,d$] State, control input, disturbance vector
  \item[$\mathcal{V}$] Lyapunov function
  \item[$\delta$] Phase angle (rad)
  \item[$\omega$] Angular frequency (rad/s)
  \item[$\Omega$] Frequency consensus variable (rad/s)
\end{IEEEdescription}}

\manuallabel{Section I}{I}\section{Introduction}
\IEEEPARstart{T}{he} introduction of Renewable Energy Sources (RES) has brought forth a new school of thought regarding the power grid structure. Typically, electricity is generated in power plants, transmitted at high voltage and distributed to the load in a centralized and unidirectional manner. However, as RES is adopted at the distribution level, power flow is becoming bidirectional. This is forming a distributed structure of small-scale generation sources such as Distributed Energy Resources (DERs). Increased intermittent RES usage, coupled with the effects of climate change, are transforming power grid operation. This transformation is affecting the reliability, robustness and resilience of the grid.

To address the grid transformation, the integration of DERs can be leveraged to form Microgrids (MGs){\color{black}. MGs are} a microcosm of the bulk power grid with local sources and loads. Using suitable control schemes, MGs provide a solution for DER integration with stable grid operation \cite{b1}. One school of thought proposes further leveraging the advantages of MGs by forming Networks of MGs (NMGs){\color{black}. NMGs can be considered as an} agglomeration of cyber-physically linked MGs that operate in a coordinated manner. This concept falls under the IEEE 2030 standards definition of smart grid interoperability \cite{b2}. {\color{black}Practical implementations include Puerto Rico's NMG-like structure with large-scale DER integration} \cite{b3}.

NMGs can operate in grid-connected or islanded modes{\color{black}. This contributes to} safely integrating RES into a large-scale power grid with scalable and resilient operation, especially during extreme scenarios such as blackouts. Hence, NMGs may be considered as a potential catalyst for added robustness, resilience and efficiency \cite{b1}. That said, {\color{black}islanded NMG operation} is still in the research and development stages. 


The operational challenges of NMGs in islanded mode affect robustness and flexibility. Such challenges have necessitated changes to the {\color{black}existing} hierarchical control scheme{\color{black}.} Hierarchical control refers to a three-layer control scheme at separate timescales{\color{black}:} frequency and voltage stabilization and power sharing at the primary level (within seconds), frequency and voltage regulation at the secondary level (within seconds to minutes) and optimization or energy management at the tertiary level (within minutes to hours). However{\color{black}, hierarchical control} was designed for a centralized, unidirectional, slow-varying and high-inertia bulk power grid. {\color{black}It may not effectively} manage the perturbations from generation and load changes{\color{black}, as well as DER and MG} connections and disconnections.

Since our objective is robust {\color{black}distributed control of NMG} frequency and voltage{\color{black}, we} focus on secondary level control in islanded mode. Primary level control {\color{black}is decentralized} and based on the well-established droop control technique. {\color{black}It defines} a linear relation between frequency (voltage) and active (reactive) power for proportional power sharing \cite{b1,b4,b5,b6}. There is a growing interest in secondary level control of NMGs, especially under different scenarios, topologies and restrictions. This can be addressed by scaling the MG schemes proposed in the literature to be used within NMGs \cite{b4,b5,b6}. 

The aforementioned issues require robust control schemes that consider the sensitivity to imbalances between local generation and load fluctuations, and cyber-physical topological uncertainties within NMGs \cite{b7}. The UNIFI consortium \cite{b8} {proposes a framework for {\color{black}Grid Forming (GFM)} Inverter-Based Resources (IBRs) {\color{black}with} fast frequency and voltage response{\color{black},} to survive disturbances while prioritizing system-wide stability. Moreover, there is a need to develop flexible distributed control to maintain supply and demand balance during prolonged outages \cite{b9}. In view of that, we will consider {\color{black}the possibility of establishing} a distributed frequency and voltage control scheme that guarantees NMG-wide robustness with respect to: (1) fluctuations in generation and load at the DER level and (2) topological uncertainties at the NMG cyber-physical level.

Before addressing secondary level control within NMGs, it is helpful to describe some of the spatiotemporal challenges facing the hierarchical control of MGs. Dörfler {\it et al.} \cite{b10} discussed the potential of real-time hierarchical control to take advantage of fast-paced IBRs. Nonetheless, {\color{black}spatially, real-time centralized} secondary level control schemes require extensive {\color{black}communication architectures and computational efforts}. {\color{black}To alleviate such requirements, distributed control schemes can be used instead}. 

Based on Nawaz \textit{et al.}'s \cite{b11} survey on secondary level control of MGs, the most studied {\color{black}distributed} secondary level control technique is consensus-based control of DERs within MGs. {\color{black}Consensus-based control is} a form of cooperative control where agents (i.e., DERs) reach an agreement on control signals to regulate frequency and voltage in an equitable manner. {\color{black}It} is suitable for influencing global behavior using distributed local controllers operating within a short timescale. A popular consensus-based approach at the secondary level is Distributed-Averaging Proportional-Integral (DAPI) control of DERs within MGs. Simpson-Porco \textit{et al.} \cite{b12} propose distributed real-time secondary level regulation, while maintaining power sharing of droop-controlled DERs. {\color{black}DAPI control} was a step towards efficient real-time control{\color{black}. It} partially recreates the performance of central controllers, such as Automatic Generation Control (AGC), in the aggregate. {\color{black}Regarding NMGs, some distributed secondary control schemes \cite{b13,b14} address grid-connected NMGs or utilize more conventional control schemes with timescale separation. One issue with the DAPI and distributed consensus control is the lack of constraints on some consensus dynamics.

Improving the consensus dynamics} might require schemes like $H_\infty$ robust control to deal with cyber-physical perturbations, per Bevrani \cite{b15}. {\color{black}One example of dealing with such perturbations is a} cyber-resilient AGC for MGs that is robust against False Data Injections (FDIs) \cite{b16}. Nonetheless, the wider range of operational and structural perturbations, {\color{black}such as load changes and MG connections and disconnections, is typically unaddressed.}

To {\color{black}partially} resolve this issue, Kammer and Karimi \cite{b17} introduced robust DAPI control with {\color{black}load disturbance rejection. In subsequent works, robustness to physical changes was addressed \cite{b18,b19} without DAPI} control. \color{black}Maaruf \textit{et al.} proposed robust distributed secondary control using feedback linearization \cite{b20,b21}. However, robustness against structural perturbations in NMG settings is yet to be explored.} Strehle \textit{et al.} \cite{b22} use a decentralized passivity-based approach at the subsystem (generation/load) level to guarantee stability in the face of load and topology changes. They also utilize Equilibrium-Independent Passivity (EIP) for stability guarantees without explicit knowledge of equilibrium points. Regardless, their work did not focus on secondary level frequency and voltage regulation. A more holistic view of robustness against operational and structural perturbations across all subsystems might be necessary for frequency and voltage regulation, while considering power sharing among DERs. Lastly, the work of Cao \textit{et al.} \cite{b23} with $H_\infty$ robust control addresses cyber-physical disruptions resulting in secondary level distributed control - but only on the communication side. 

To achieve stability guarantees against both operational and structural perturbations, Šiljak introduced the notion of connective stability \cite{b24}{\color{black}. It} ensures the stability of complex systems such as NMGs for all possible topologies or interconnections. Its importance lies in the fact that generally, there is an inversely proportional relation between the stability and connectivity of complex systems, as more interconnections will lead to more stochasticity \cite{b25}. Šiljak \cite{b26} and Ćalović \cite{b27} proposed an altered AGC scheme to enhance stability margins. Marinovici \textit{et al.} \cite{b28} established a connectively stable decentralized frequency control scheme for the same purpose at the secondary level. Both works were limited to conventional large-scale generation power grids. To address this issue, Khanbaghi and Zečević \cite{b29,b30} proposed a robust multi-layer control strategy utilizing the concept of connective stability for robust energy management, since an NMG is only stable if the MGs that compose it are stable. The approach {\color{black}is a Linear Matrix Inequality (LMI)-based strategy that} guarantees stability in the face of topological uncertainties, but it focuses exclusively on tertiary level control. Other layers in the hierarchical control scheme are yet to be explored.

\begin{figure*}[!b]\centering\manuallabel{Fig. 2.1}{1}\vspace{-12pt}
  \includegraphics[width=\textwidth]{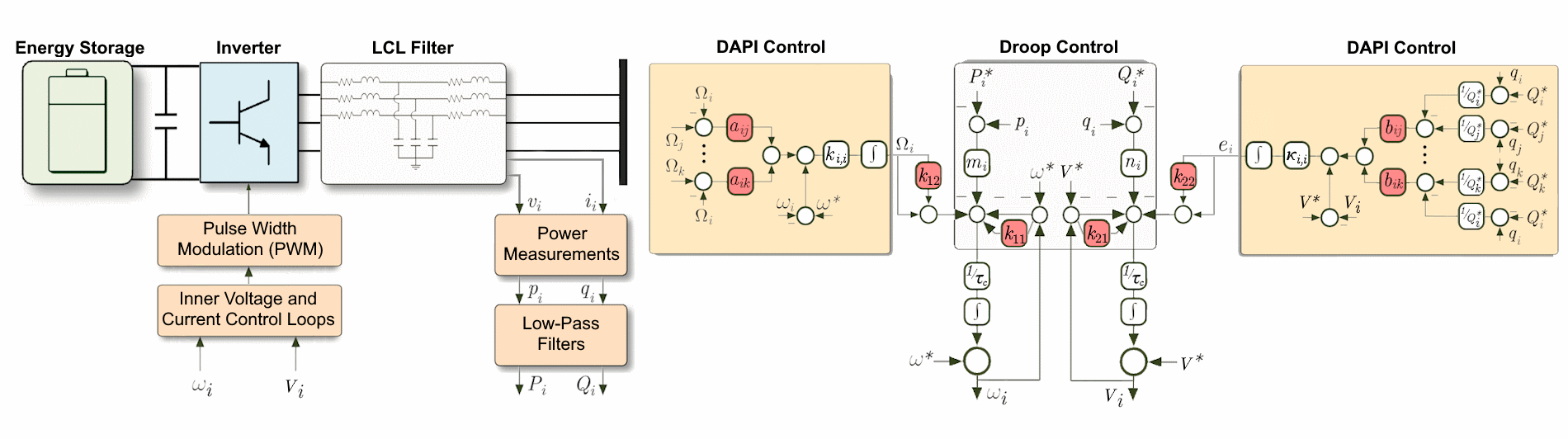}
  \caption{\color{black}The Distributed Energy Resource (DER) block diagram with primary level and secondary level control. The red blocks indicate the parameters computed by the proposed control strategy.}
\end{figure*}

The literature surveyed in this section indicates that DAPI-based distributed secondary level control is an effective method for real-time frequency and voltage control of droop-controlled MGs. However, it also suggests {\color{black}that} a more holistic view of MG secondary level control is needed. {\color{black}Performance guarantees during fluctuations from local imbalances must be considered, as well as cyber-physical disturbances from neighboring DERs or MGs within an NMG. \textit{The contribution of this work is to develop a robust distributed secondary level control strategy for NMGs}, and the process is summarized as follows}: 
\begin{enumerate}
    \item {\color{black}Establishing} a modified state-space model representing DAPI-controlled inverters with consensus dynamics as uncertainties as well as added state feedback, {\color{black}for an additional degree of freedom that ensures robustness.}
    \item {\color{black}Applying} the concept of connective stability of interconnected systems to NMGs, to provide distributed secondary level frequency and voltage control that guarantees stability and ensures robustness to operational and structural perturbations.
    \item {\color{black}Using} the invariant ellipsoid technique to {\color{black}bound} the effects of the disturbances and uncertainties within the NMG{\color{black}, for more robust operation.}
    \item {\color{black}Formulating} the control design as an LMI optimization problem, which combines the aforementioned uses of connective stability and the invariant ellipsoid technique to achieve the desired stability and guarantee robustness.
\end{enumerate}
We validate the proposed strategy by simulating the dynamic behavior of an NMG consisting of 3 MGs and 5 DERs using MATLAB\textsuperscript{\textregistered}/Simulink\textsuperscript{\textregistered}. The proposed scheme was tested in various scenarios based on cyber-physical perturbations. 

The remainder of this paper is organized as follows: Section \ref{Section II} introduces the DER and NMG modeling, followed by the robust stability conditions in Section \ref{Section III} and the derivation of the proposed control scheme in Section \ref{Section IV}. Simulation results are shown in Section \ref{Section V} before listing our concluding remarks and future work in Section \ref{Section VI}.

\manuallabel{Section II}{II}\section{Distributed Energy Resource and Network of Microgrids Modeling}

In this section, we introduce a DER model for a droop-controlled inverter with DAPI frequency and voltage control, aggregated into an NMG system operating in islanded mode. The DER block diagram is shown in Fig. \ref{Fig. 2.1}. Its main components are a Battery Energy Storage System (BESS), {\color{black} a droop-controlled} inverter and an LCL filter. {\color{green}It is assumed that the BESS is pre-charged by a Grid Following (GFL) generation source such as a Photovoltaic (PV) system. Since the GFL source does not directly contribute towards frequency and voltage regulation, the PV generation is not modeled.} The primary level and secondary level control {\color{black}schemes are} explained in the following subsections.

\subsection{Primary Level Control}
{\color{black}Following \cite{b31,b32,b33,b34}, we} can establish a state-space model representing the DER primary level and secondary level control dynamics using the droop control expression with Low-Pass Filter (LPF) dynamics. The {\color{black}$i^{th}$} droop-controlled inverter model is governed by the following equations{\color{black}\cite{b31,b32}}:
\begin{subequations}\begin{align}
&\label{(1a)}\dot{\delta}_i(t) = \omega_i(t) - \omega^* \\
&\label{(1b)}\omega_i(t) = \omega^*  -m_i(P_i(t)-P_i^*) \\
&\label{(1c)}V_i(t) = V^*  -n_i(Q_i(t)-Q_i^*)\end{align}\end{subequations}
where {\color{black}$\delta_i(t)$ is the phase angle}, $\omega_i(t)$ and $\omega^*$ are the actual and setpoint frequencies, respectively, $m_i$ is the active power-frequency droop gain and $P_i(t)$ and $P_i^*$ are the actual and {\color{black}setpoint} active power outputs. In the third equation, $V_i(t)$ and $V^*$ represent the actual and setpoint voltages{\color{black},} respectively, $n_i$ is the reactive power-voltage droop gain and $Q_i(t)$ and $Q_i^*$ are the actual and {\color{black}setpoint} reactive power outputs. 

Equations \eqref{(1b)} and \eqref{(1c)} can be reformulated as differential equations representing frequency and voltage using LPF dynamics. LPFs are used to enhance power quality by attenuating higher-order frequencies and transients. The time response for the active and reactive power LPFs, assuming equal cutoff frequencies, is{\color{black}\cite{b33}}:
\begin{subequations}\begin{align}
&\label{(2a)}\tau_c\dot{P}_i(t) = -P_i(t) + p_i(t) \\
&\label{(2b)}\tau_c\dot{Q}_i(t) = -Q_i(t) + q_i(t)\end{align}\end{subequations}
where $p_i(t)$ and $q_i(t)$ are the instantaneous active and reactive powers at the Point of Common Coupling (PCC) representing the {\color{black}filters'} input signals{\color{black}.} $\tau_c$ is the filters' cutoff time constant. 

{\color{black}After taking the derivative of \eqref{(1b)} and \eqref{(1c)}, substituting in \eqref{(2a)} and \eqref{(2b)} and redefining $P_i(t),Q_i(t)$ using \eqref{(1b)} and \eqref{(1c)}, we obtain \cite{b31,b32,b33}:}
\begin{subequations}\begin{align}
&\label{(3a)}\tau_c\Delta \dot{\omega}_i(t) = -\Delta \omega_i(t) - m_i\Delta p_i(t) \\
&\label{(3b)}\tau_c\Delta \dot{V}_i(t) = -\Delta V_i(t) - n_i\Delta q_i(t)\end{align}\end{subequations}
where $\Delta \omega_i(t) = \omega_i(t) - \omega^*$, $\Delta V_i(t) = V_i(t) - V^*$, $\Delta p_i(t) = p_i(t) - P_i^*$ and $\Delta q_i(t) = q_i(t) - Q_i^*$.

It is well known that after a disturbance, the proportional droop controller introduces frequency and voltage steady-state errors. In the next subsection, we formulate the secondary level control terms to eliminate frequency steady-state errors and mitigate voltage steady-state errors using the DAPI scheme.

\subsection{Secondary Level Control}
Since droop control is a form of proportional control, any changes in inverter active and reactive power outputs would lead to frequency and voltage deviations in the form of steady-state errors{\color{black}. This requires} some type of integral control that maintains power sharing between DERs. In view of that, secondary level control {\color{black}frequency and voltage} terms $u_{\omega,i}(t), u_{V,i}(t)$ are added to compensate for the steady-state error caused by primary level control \cite{b31}:



\begin{subequations}\begin{align}
&\label{(4a)}\tau_c\Delta \dot{\omega}_i(t) = -\Delta \omega_i(t) - m_i\Delta p_i(t) + u_{\omega,i}(t) \\
&\label{(4b)}\tau_c\Delta \dot{V}_i(t) = -\Delta V_i(t) - n_i\Delta q_i(t) + u_{V,i}(t)\end{align}\end{subequations}

To deal with frequency and voltage steady-state errors while maintaining droop control power sharing, we adopt {\color{black}the DAPI control scheme \cite{b12}. DAPI control is based} on distributed-averaging consensus control. {\color{black}This avoids the single point of failure weakness in centralized AGC-based secondary level control \cite{b35}. The DAPI frequency control dynamics are defined as} \cite{b12}:
\begin{equation}\label{(5)} k_{i,i}\dot{\Omega}_i(t) = -\Delta \omega_i(t) - \sum_{j=1}^Na_{ij}(\Omega_i(t) - \Omega_j(t))\end{equation}
where $\Omega_i(t){\color{black},\Omega_j(t)}$ are the secondary level frequency control consensus variable{\color{black}s}, $k_{i,i}$ is a positive inverse DAPI integral gain{\color{black},} $a_{ij}$ is a non-negative adjacency matrix coefficient corresponding to neighboring DER or MG (coupling) links{\color{black}, $N$ is the total number of DERs and $j$ refers to neighboring DERs}. {\color{black}We also consider the following DAPI voltage controller \cite{b12}}:
\begin{equation}\label{(6)}\kappa_{i,i}\dot{e}_i(t) = -\xi_i\Delta V_i(t) - \sum_{j=1}^Nb_{ij}\left(\frac{Q_i(t)}{Q_i^*} - \frac{Q_j(t)}{Q_j^*}\right)\end{equation}
where $e_i(t)$ is the secondary level voltage control consensus variable, $\kappa_{i,i}$ is a positive inverse DAPI integral gain{\color{black}, $\xi_i$ is a non-negative voltage deviation gain and $b_{ij}$ is a} non-negative adjacency matrix coefficient corresponding to neighboring DER or MG (coupling) links. 

{\color{black}We note that} \eqref{(6)} contains disturbance variables $Q_i(t)$ (and $Q_j(t)$), which are defined differently compared to $\Delta q_i(t)$ (or $\Delta q_j(t)$) in \eqref{(3b)}. To address this issue, we will add $-Q_i^*/Q_i^*$ and $Q_j^*/Q_j^*$, and use instantaneous reactive power values $q_i(t)$ instead of filtered ones $Q_i(t)$. {\color{black}Accordingly, we obtain}:
\begin{equation}\label{(7)}\kappa_{i,i}\dot{e}_i(t) = -\xi_i\Delta V_i(t) - \sum_{j=1}^Nb_{ij}\left(\frac{\Delta q_i(t)}{Q_i^*} - \frac{\Delta q_j(t)}{Q_j^*}\right)\end{equation}
If $Q^*_i$ or $Q^*_j$ are not taken to be the inverter rating, then $b_{ij}$, $Q_i^{*^{-1}}$ and $Q_j^{*^{-1}}$ in \eqref{(7)} are taken to be $0$ if $Q_i^*=0$ or $Q_j^*=0$. The effects of each term in \eqref{(5)}, \eqref{(7)} in terms of regulation and power sharing are explained in {\color{black}\cite{b12,b36}}.  


To provide a degree of freedom for control purposes, we add $\Delta u_{\omega,i}(t)$ and $\Delta u_{V,i}(t)$ as frequency and voltage control terms to be determined by state feedback control of the DAPI-controlled $i^{th}$ DER state-space model:
\begin{subequations}\begin{align}
&\label{(8a)}u_{\omega,i}(t) = \Omega_i(t) + \Delta u_{\omega,i}(t) \\
&\label{(8b)}u_{V,i}(t) = e_i(t) + \Delta u_{V,i}(t)\end{align}\end{subequations}

In the following subsection, we formulate our $i^{th}$ DER state-space model {\color{black}describing primary level and secondary level control}.

\subsection{{\color{black}Distributed Energy Resource} State-Space Model}\vspace{-7pt}
By combining equations \eqref{(4a)} with \eqref{(8a)} and \eqref{(4b)} with \eqref{(8b)}, {\color{black}we can describe a state-space model for primary and secondary level control of droop-controlled DERs with DAPI control.} The state-space model of the DAPI-controlled $i^{th}$ DER can be represented in the general form:\pagebreak
\begin{equation}\label{(9)}\begin{aligned} \dot{x}_i & = A_ix_i + B_iu_i + E_id_i + h_i(x) + g_i(d) \\ & = A_ix_i + B_iu_i + E_id_i + \\ & \ \ \ \underbrace{\sum_{j=1}^N(\Delta A_{ii}x_i + \Delta A_{ij}x_j)}_{h_i(x)} + \underbrace{\sum_{j=1}^N(\Delta E_{ii}d_i + \Delta E_{ij}d_j)}_{g_i(d)}\end{aligned}\end{equation}
In \eqref{(9)}, {\color{black}state $x_i$, input $u_i$ and disturbance $d_i$ vectors} are defined as $x_i = \begin{bmatrix}\Delta\omega_i & \Omega_i & \Delta V_i & e_i\end{bmatrix}^T$, $u_i = \begin{bmatrix}\Delta u_{\omega,i} & \Delta u_{V,i}\end{bmatrix}^T$ and $d_i = \begin{bmatrix}\Delta p_i & \Delta q_i\end{bmatrix}^T$. {\color{black}$A_i,B_i$ and $E_i$ are the state, input and disturbance matrices.} The terms $\Delta A_{ii}, \Delta E_{ii}, \Delta A_{ij}$ and $\Delta E_{ij}$ in this equation represent the {\color{black}state and disturbance} uncertain consensus dynamics interactions with neighboring DERs at any point in time. $h_i(x)$ and $g_i(d)$ are the state and disturbance consensus terms{\color{black},} respectively. The expanded matrix-vector form of \eqref{(9)} is shown in \eqref{(10)}.

\begin{table*}[!b]\vspace{-8pt}\noindent\rule{\textwidth}{\arrayrulewidth} \vspace{8pt}\begin{equation}\label{(10)}\begin{split}\footnotesize\underbrace{\begin{bmatrix}\Delta \dot{\omega}_i \\ \dot{\Omega}_i\\\Delta \dot{V}_i \\ \dot{e}_i\end{bmatrix}}_{\dot{x}_i} = 
\underbrace{\begin{bNiceArray}{c c:c c}-\frac{1}{\tau_c} & \frac{1}{\tau_c}  & 0 & 0 \\ -\frac{1}{k_{i,i}} & 0 & 0 & 0 \\ \hdottedline 0 & 0 &  -\frac{1}{\tau_c} & \frac{1}{\tau_c} \\ 0 & 0 &- \frac{\xi_i}{\kappa_{i,i}} & 0\end{bNiceArray}}_{A_i}\underbrace{\begin{bmatrix}\Delta \omega_i \\ \Omega_i \\ \Delta V_i \\ e_i\end{bmatrix}}_{x_i} + 
\underbrace{\begin{bNiceArray}{c:c}\frac{1}{\tau_c} & 0 \\ 0 & 0 \\ \hdottedline 0 & \frac{1}{\tau_c} \\ 0 & 0\end{bNiceArray}}_{B_i}\underbrace{\begin{bmatrix}\Delta u_{\omega,i} \\ \Delta u_{V,i}\end{bmatrix}}_{u_i} +
\underbrace{\begin{bNiceArray}{c:c}-\frac{m_i}{\tau_c} & 0 \\ 0 & 0 \\ \hdottedline 0 & -\frac{n_i}{\tau_c} \\ 0 & 0\end{bNiceArray}}_{E_i}\underbrace{\begin{bmatrix}\Delta p_i \\ \Delta q_i\end{bmatrix}}_{d_i} \\
\footnotesize+ \sum_{j=1}^N\left(\underbrace{\begin{bNiceArray}{c c:c c}0 & 0 & 0 & 0 \\ 0 & -\frac{a_{ij}}{k_{i,i}} & 0 & 0 \\ \hdottedline 0 & 0 &  0 & 0 \\ 0 & 0 & 0 & 0\end{bNiceArray}}_{\Delta A_{ii}}\underbrace{\begin{bmatrix}\Delta \omega_i \\ \Omega_i \\ \Delta V_i \\ e_i\end{bmatrix}}_{x_i} + 
\underbrace{\begin{bNiceArray}{c c:c c}0 & 0  & 0 & 0 \\ 0 & \frac{a_{ij}}{k_{i,i}} & 0 & 0 \\ \hdottedline 0 & 0 & 0 & 0 \\ 0 & 0 & 0 & 0\end{bNiceArray}}_{\Delta A_{ij}}\underbrace{\begin{bmatrix}\Delta \omega_j \\ \Omega_j \\ \Delta V_j \\ e_j\end{bmatrix}}_{x_j}\right)
\footnotesize+ \sum_{j=1}^N\left(\underbrace{\begin{bNiceArray}{c:c}0 & 0 \\ 0 & 0 \\ \hdottedline 0 & 0 \\ 0 & -\frac{b_{ij}}{\kappa_{i,i}Q_i^*}\end{bNiceArray}}_{\Delta E_{ii}}\underbrace{\begin{bmatrix}\Delta p_i \\ \Delta q_i\end{bmatrix}}_{d_i} + 
\underbrace{\begin{bNiceArray}{c:c}0 & 0 \\ 0 & 0 \\ \hdottedline 0 & 0 \\ 0 & \frac{b_{ij}}{\kappa_{i,i}Q_j^*}\end{bNiceArray}}_{\Delta E_{ij}}\underbrace{\begin{bmatrix}\Delta p_j \\ \Delta q_j\end{bmatrix}}_{d_j}\right)\end{split}\normalsize\end{equation}\nopagebreak\end{table*}    

Such a model must adhere to a number of system constraints. Typically, exogenous disturbances such as $d_i$ are bounded, and we can treat them as smooth functions that satisfy:
\begin{equation}\label{(11)}d_i^T\bar{S}_id_i \leq 1\end{equation}
where $\bar{S}_i = \frac{1}{\Delta\bar{s}_i^2}I_{2\times2}$, and $\Delta \bar{s}_i$ represents the maximum apparent power deviation of each inverter.

\vspace{5.5pt}
\manuallabel{Remark 1}{1}\textit{Remark 1.}  Equation \eqref{(11)} is a reflection of the inverter capacity, which can be normalized. This allows us to set the right-hand side to 1{\color{black}, for a more convenient LMI formulation.}
\vspace{5.5pt}



{\color{black}Lastly,} in the following, we will consider state feedback control of the form $u_i = K_ix_i${\color{black}, with state feedback gain matrix $K_i$}. To maintain consensus and power sharing, matrix $K_i$ is assumed to be identical for all inverters $i \ \in\ \{1,2,\dots, N\}$. We can therefore represent each such a matrix as:
{\color{black}\begin{equation*}K_i = \begin{bmatrix}k_{11} & k_{12} & 0 & 0 \\ 0 & 0 & k_{21} & k_{22}\end{bmatrix}\end{equation*}
where $k_{11},k_{12},k_{21},k_{22}$ are }{\color{black}the state feedback gains.}

We now proceed to describe how the DER state-space model can be used to obtain an aggregate NMG model for the LMI optimization problem.

\subsection{{\color{black}Network of Microgrids} Model}
{\color{black}Per Simpson-Porco \textit{et al.}'s \cite{b12} work, DAPI control does not require explicit knowledge of the power system topology, impedance or loads. Thus, the physical model does not need to be explicitly considered in the NMG state-space model. The DER output power terms suffice, along with their communication links. This allows us to establish a primary and secondary level control-based model that implicitly accounts for connective stability and cyber-physical disturbances.} 

{\color{black}The NMG model is a composite (aggregate) model of all DERs within the network. We assume each MG is a miniature power system that can operate autonomously in islanded mode or in grid-connected mode. We focus our attention on NMGs because they are more likely to experience operational and structural perturbations compared to MGs. Another important point is {\color{black}that} the control strategy requires a holistic view of the NMG model, but it only needs to be computed once offline. Accordingly, an} aggregated expression for the state-space model of an NMG with $N$ interconnected DERs {\color{black}within MGs} has the following general form:
\begin{equation}\label{(13)}\begin{aligned}\dot{x} & = A_Dx + B_Du + E_Dd + h_D(x) + g_D(d) \\ & = (A_D + \Delta A_D)x + B_Du + (E_D + \Delta E_D)d\end{aligned}\end{equation}
where $A_D, B_D$ and $E_D$ are block {\color{black}state, input and disturbance} matrices, and $x = \begin{bmatrix}x_1 & x_2 & \dots & x_N\end{bmatrix}^T$, $u = \begin{bmatrix}u_1 & u_2 & \dots & u_N\end{bmatrix}^T$ and $d = \begin{bmatrix}d_1 & d_2 & \dots & d_N\end{bmatrix}^T$ are the aggregated states, inputs and disturbances. The terms $h_D = \begin{bmatrix}h_1 & h_2 & \dots & h_N\end{bmatrix}^T$ and $g_D = \begin{bmatrix}g_1 & g_2 & \dots & g_N\end{bmatrix}^T$ represent the state consensus and disturbance consensus dynamics, and $\Delta A_D$ and $\Delta E_D$ {\color{black}are the composite state and disturbance consensus dynamics interactions matrices.}

The overall feedback control has the form:
\begin{equation}\label{(14)}u = K_Dx\end{equation}
{\color{black}for composite gain matrix} $K_D = diag\{K_1,K_2,\dots,K_N\}$. The composite model bounds can be expressed in the following aggregated form \cite{b30}:
\begin{equation}\label{(15)}d^T\bar{S}_Dd \leq 1\end{equation}
\vspace{-6mm}
\begin{subequations}\begin{align}
\label{(16a)}&h_D(x)^Th_D(x) \leq \alpha^2 x^TH^THx \\
\label{(16b)}&g_D(d)^Tg_D(d) \leq \beta^2 d^TG^TGd\end{align}\end{subequations}
for block diagonal matrices of normalized {\color{black}maximum apparent power deviation $\bar{S}_D$, state consensus connective strength $\alpha$ and bound matrix $H$ and disturbance consensus connective strength $\beta$ and bound matrix $G$.}

The system dynamics described in \eqref{(10)} can be represented in compact form as{\color{black}\cite{b37}}:
\begin{subequations}\begin{align}
\label{(17a)}&\tau_c\Delta\dot{\omega}  = -\Delta \omega + \Omega + \Delta u_\omega -M_D \Delta p \\ 
\label{(17b)}&k_i\dot{\Omega} = -\Delta \omega - \mathcal{L}_{a}\Omega \\ 
\label{(17c)}&\tau_c\Delta\dot{V}  = -\Delta V + e + \Delta u_V -N_D \Delta q \\ 
\label{(17d)}&\kappa_i\dot{e} = -\xi\Delta V - \mathcal{L}_{b}Q^{*^{-1}}\Delta q\end{align}\end{subequations}
{\color{black}where $\tau_c,k_i,\kappa_i$ and $\xi$ are composite matrices of the filters' time constants, inverse frequency and voltage consensus integral gains and voltage deviation gains, with $M_D,N_D$ and $Q^{*^{-1}}$ being diagonal frequency droop gain, voltage droop gain and inverse reactive power setpoint $N \times N$} matrices. $\mathcal{L}_{a}$ and $\mathcal{L}_{b}$ are graph Laplacian matrices that reflect the structure of the NMG's underlying communication links. Further context on graph theoretic concepts can be found in \cite{b37}.

\vspace{5.5pt}
\manuallabel{Remark 2}{2}\textit{Remark 2.} To simplify the modeling and simulation processes, $\tau_c$, $k_i$, $\kappa_i$ and $\xi$ are taken to be identical for all inverters. In general, each inverter can have {\color{black}different} $\tau_c$, $k_i$, $\kappa_i$ and $\xi$ values depending on the NMG setup.
\vspace{5.5pt}

The described system is stable but not necessarily asymptotically stable. One of the reasons for this is that $\Delta V$ can be non-zero. {\color{black}In addition, $\mathcal{L}_{a}$ and $\mathcal{L}_{b}$ are guaranteed to be singular,} which typically leads to non-zero consensus variables $\Omega$ and $e$. When it comes to consensus dynamics, LaSalle's invariance principle ensures that the system trajectory converges towards the largest invariant or equilibrium set $\mathcal{D} \subset \mathbb{R}^{4N}$. This can be formally expressed as \cite{b37}:
\begin{equation}\label{(18)}\lim_{t \rightarrow \infty}\inf_{x_e \in \mathcal{D}}||x - x_e||_2 = 0\end{equation}
{\color{black}where $x_e$ is the equilibrium point.} In the disagreement space, the difference between consensus variables generally converges to zero as $t \rightarrow \infty$. However, this doesn't necessarily apply to voltage consensus $e$ \cite{b38}. In view of that, we must first consider general stability conditions for such systems. Once we have specified them, we can proceed to develop a suitable control strategy.

\manuallabel{Section III}{III}\section{Practical (Robust) Stability Conditions}
{\color{black}In this section, we'll consider the conditions upon which stability is guaranteed for the system described in state-space form \eqref{(13)}. The system} can be asymptotically stable, but this is not always guaranteed. It is clear that the system is asymptotically stable irrespective of graph topology if $u = d = 0$. However, the presence of persistent disturbances and uncertainties means the standard theoretical techniques for establishing asymptotic stability may not be applicable. With that in mind, we {\color{black}utilize} practical stability, which is commonly used when dealing with real-world problems. Practical stability was introduced by LaSalle and Lefschetz \cite{b39}, and is less stringent than Lyapunov stability since it refers to a system's ability to stay within acceptable bounds \cite{b40}. To limit the sensitivity of the system states to the magnitude of forced responses, we can invoke Input-to-State Stability (ISS) theory and its dissipation property. 

For state-space models, Sontag \cite{b41,b42} introduced the concept of ISS and concluded that it implies two things: (a) the unforced system must be asymptotically stable and (b) the state trajectory of the forced system must asymptotically approach a region around the origin. 
The ISS property is defined in terms of a decay estimate of solutions $x(t)$, and is equivalent to the validity of a dissipation inequality holding along all possible trajectories for a suitable Lyapunov function {\color{black}$\mathcal{V}(x)$} defined as: 
\begin{equation}\label{(19)}\mathcal{V}(x)=x^T\boldsymbol{P}_Dx\end{equation} 
{\color{black}where} $\boldsymbol{P}_D$ is a symmetric positive definite matrix such that the derivative $\dot{\mathcal{V}}(x)$ must satisfy the following supply rate condition{\color{black}\cite{b43,b44}}:

\begin{equation}\label{(20)}\dot{\mathcal{V}}(x) \leq \begin{bmatrix}d & x\end{bmatrix}\begin{bmatrix}\mathbf{Q}+\Delta \mathbf{Q} & \mathbf{R}+\Delta \mathbf{R} \\ (\mathbf{R} + \Delta \mathbf{R})^T & -\mathbf{S} + \Delta \mathbf{S}\end{bmatrix}\begin{bmatrix}d \\ x\end{bmatrix}\end{equation}
Expression \eqref{(20)} implies that the system consumes or conserves (but doesn't generate) net energy{\color{black},} with respect to the supply rate with derived respective weighting supply rate matrices $\mathbf{Q}, \mathbf{S},\mathbf{R}$ and uncertainties $\Delta\mathbf{Q},\Delta\mathbf{S}$ and $\Delta\mathbf{R}$. {\color{black}It} can be shown that the system in \eqref{(13)} is dissipative with respect to $d(t)$ and $x(t)$ with a Lyapunov function $\mathcal{V}(x)$ if:
\begin{equation}\label{(21)}\dot{\mathcal{V}}(x) \leq -x^T\mathbf{S}x  + d^T{\color{black}(\mathbf{Q}+\Delta\mathbf{Q})}d + x^T\Delta \mathbf{S} x + 2d^T(\mathbf{R} + \Delta \mathbf{R})x\end{equation}
(The proof is provided in Appendix \ref{Appendix A}).

Equation \eqref{(21)} ensures stability regardless of equilibrium point ${\color{black}x_e} \in \mathcal{D}$. {\color{black}To analyze stability conditions, we consider} equilibrium-independent dissipativity. According to \cite{b43,b44}, equilibrium-independent dissipativity can guarantee the dissipativity condition referenced to two system trajectories, one of which is a fixed point, i.e., any equilibrium point ${\color{black}x_e} \in \mathcal{D}$.

Given Lyapunov function $\mathcal{V}(\tilde{x}) = \tilde{x}^T\boldsymbol{P}_D\tilde{x}$, an arbitrary equilibrium point ${\color{black}x_e}$ and a disturbance ${\color{black}d_e}$, inequality \eqref{(20)} implies that the system in \eqref{(13)} is equilibrium-independent dissipative if:

\begin{equation}\label{(22)}\dot{\mathcal{V}}(\tilde{x}) \leq  -\tilde{x}^T\mathbf{S}\tilde{x} + \tilde{d}^T{\color{black}(\mathbf{Q}+\Delta\mathbf{Q})}\tilde{d} + \tilde{x}^T\Delta \mathbf{S} \tilde{x} + 2\tilde{d}^T(\mathbf{R} + \Delta \mathbf{R})\tilde{x}
\end{equation}
where $\tilde{x}=x-x_e$ and $\tilde{d}=d-d_e$ {\color{black}such that $d_e$ is the disturbance at equilibrium point $x_e$. Under the associated invariance conditions, for disturbance $d=d_e$ at a given equilibrium point, the equilibrium is Lyapunov stable if $\mathbf{S}-\Delta \mathbf{S} \geq 0$}. As we already noted, the system converges towards an equilibrium contained in the set $\mathcal{D}$ (according to LaSalle's invariance principle).


Since the system in \eqref{(13)} is subjected to both exogenous disturbances and matrix uncertainties, it is a suitable candidate for practical stability techniques. Per Polyak \textit{et al.} \cite{b45,b46}, this involves approximating the reachable set of the system. The presence of persistent exogenous disturbances and uncertainties implies that the system is unlikely to settle {\color{black}at an equilibrium} and remain there for all $t \geq 0$ \cite{b39,b40}. 

By constraining the reachable set, we can achieve our goal of limiting the system trajectory sensitivity to the magnitude of forced responses. To accomplish that, we can utilize connective stability and invariant ellipsoid techniques, both of which involve a formulation of Lyapunov stability conditions that is similar to what was discussed in this section. It is worth noting that the invariant ellipsoid technique can be used for initial set containment to account for the set of equilibria $\mathcal{D}$ and a given range on states $x$. This is defined in the next section. 

Using the state-space model in \eqref{(13)}, the connective stability and invariant ellipsoid {\color{black}techniques} will be defined and adopted to compute $\Delta u_{\omega}, \Delta u_{V}$ using state feedback control, along with the connective strength-related coupling terms $a_{ij}$ and $b_{ij}$. Our goal will be to ensure stability in the presence of both operational and structural perturbations within NMGs.

\manuallabel{Section IV}{IV}\section{Robust and Connectively Stable Linear Matrix Inequality-Based Strategy}
In this section, we establish the connective stability conditions based on the work of Šiljak in \cite{b24,b47}. This will allow us to formulate an LMI-based problem that guarantees stability when the system is subjected to structural perturbations. Additionally, we will employ the invariant ellipsoid technique based on the work of Polyak in \cite{b45,b46}. These studies deal with disturbance rejection and increased robustness to perturbations for the same class of systems.

\subsection{Connective Stability}
As explained in Section \ref{Section I}, connective stability concerns the guaranteed stability of complex systems (NMGs) for all possible topologies or interconnections \cite{b24}. After dropping explicit references to time, the DAPI coupling terms $a_{ij}$ and $b_{ij}$ can be redefined {\color{black}to determine} connective stability, using the connective strength bounding terms $\alpha$ and $\beta$:
\begin{equation}\label{(23)}(a_{ij},b_{ij}) = (\alpha \boldsymbol{e}_{ij},\beta \boldsymbol{e}_{ij})\end{equation}
In \eqref{(23)}, $\boldsymbol{e}_{ij}$ is the connective strength, with a minimum of 0 and a maximum of $\boldsymbol{e}_{ij} = 1$. This allows us to define the ${\color{black}N \times N}$ fundamental interconnection matrix $\bar{\boldsymbol{E}} = (\boldsymbol{\bar{e}}_{ij})$ {\color{black}(with fundamental connective strength $\boldsymbol{\bar{e}}_{ij}$)}, which represents the communication links between DERs that were determined in the NMG design phase\cite{b26}:
\begin{equation}\label{(24)}\boldsymbol{\bar{e}}_{ij} = \begin{cases} 
      1, & \hspace{-4.5pt} \text{$x_j,d_j$ {\bf can} occur in $\sum {\color{black}\Delta A_{ij}}x_j, \sum {\color{black}\Delta E_{ij}}d_j$,} \\
      0, & \hspace{-4.5pt} \text{$x_j,d_j$ {\bf can't} occur in $\sum {\color{black}\Delta A_{ij}}x_j, \sum {\color{black}\Delta E_{ij}}d_j$.}
   \end{cases}\end{equation}
A communication link between two DERs is assumed to exist only if there is a direct or indirect physical link between DERs or MGs. Consequently, $\bar{\boldsymbol{E}}$ describes the possible connection status between DERs in the NMG design phase. To describe the connection status during the NMG operation phase, the interconnection matrix $\boldsymbol{E}$ is utilized{\color{black}\cite{b26}}: 
\begin{equation}\label{(25)}\boldsymbol{e}_{ij} = \begin{cases} 
      1, & \hspace{-4.5pt}  \text{$x_j,d_j$ {\bf do} occur in $\sum {\color{black}\Delta A_{ij}}x_j, \sum {\color{black}\Delta E_{ij}}d_j$,} \\
      0, & \hspace{-4.5pt} \text{$x_j,d_j$ {\bf don't} occur in $\sum {\color{black}\Delta A_{ij}}x_j, \sum {\color{black}\Delta E_{ij}}d_j$.}
   \end{cases}\end{equation}
The elements $\boldsymbol{e}_{ij}$ represent the connective strength between the $i^{th}$ DER and its $j^{th}$ neighbor at any given time $t$. The matrix $\boldsymbol{E} = (\boldsymbol{e}_{ij})$ with elements $\boldsymbol{e}_{ij} \in [0,1]$, is said to be generated by the fundamental interconnection matrix $\bar{\boldsymbol{E}} = (\boldsymbol{\bar{e}}_{ij})$ if $\boldsymbol{\bar{e}}_{ij} = 0$ implies $\boldsymbol{e}_{ij}(t) = 0$. We will say that the system in \eqref{(13)} is connectively stable if it is stable in the sense of Lyapunov for all $\boldsymbol{E}$ that are generated by $\bar{\boldsymbol{E}}$ \cite{b24}.

In essence, connective stability guarantees robustness to topological uncertainties within NMGs. In Simpson-Porco \textit{et al.}'s DAPI control scheme \cite{b12,b48}, $a_{ij}$ and $b_{ij}$ can be manually tuned to improve transient behavior. Bearing that in mind, we propose to systematically tune $a_{ij}$ and $b_{ij}$ using parameters $\alpha$ and $\beta$, to ensure robustness to structural perturbations such as disconnections or reconnections. 

To accomplish this, an LMI-based strategy will be formulated, which guarantees robust connective stability of the NMG by treating the adjacency coupling term as an additive perturbation. This corresponds to setting $h_D(x) = \Delta A_Dx$ and $g_D(d) = \Delta E_Dd$ in the composite NMG state-space model, where $\Delta A_Dx,\Delta E_Dd$ are bounded as in \eqref{(16a)} and \eqref{(16b)}. This will allow us to compute suitable connective strength bounds, so that the $i^{th}$ DER is robust to NMG structural perturbations.

\manuallabel{Section IV-B}{IV-B}\subsection{Proposed Control Strategy Formulation}
Our objective in this subsection is to show how the LMI approach can be applied to the NMG model described in \eqref{(13)}. {\color{black}For stability guarantees against} perturbations, we will focus on computing appropriate bounds on parameters $\alpha$ and $\beta$, and on constructing a decentralized gain matrix $K_D$. Following \cite{b24} and \cite{b45,b46,b47}, we will utilize the Lyapunov function in \eqref{(19)}, where the matrix $\boldsymbol{P}_D$ is assumed to be positive definite and block diagonal. This Lyapunov function will have to satisfy the dissipativity condition defined by inequality \eqref{(21)}.

As we noted earlier, {\color{black}we aim} to bound the trajectory $x(t)$ (and by extension the reachable set $\mathcal{R}$) for all admissible $d$, $\Delta A_D$ and $\Delta E_D$. {\color{black}This} problem is usually solved using $H_\infty$ control by minimizing the gain to achieve effective disturbance rejection. However, typically, the reachable sets considered assume zero initial conditions to isolate the forced response of inputs and disturbances from the system's free internal response. This is not always a valid assumption for the model in $\eqref{(13)}$, where voltage {\color{black}deviations and consensus can converge toward non-zero steady-state values. To approximate the reachable set $\mathcal{R}$ while accounting for {\color{black}bounded} non-zero initial conditions, the invariant ellipsoids technique} can be used \cite{b45,b46}. This entails defining {\color{black}an ellipsoid $\mathcal{E}_x$ as \cite{b45,b46}}: 
\begin{equation}\label{(26)}\mathcal{E}_x = \{x \in \mathbb{R}^{4N}: x^TY_D^{-1}x \leq 1\ |\ Y_D = Y_D^T > 0\}\end{equation}
where $Y_D^{-1} = \kappa_Y^{-1}\boldsymbol{P}_D$ and $Y_D \geq Y_{D,0}$ for initial set containment matrix $Y_{D,0}$. This condition is equivalent to:
\begin{equation}\label{(27)}\mathcal{E}_x = \{x \in \mathbb{R}^{4N}: x^T\boldsymbol{P}_Dx \leq 1\ |\ \boldsymbol{P}_D = \boldsymbol{P}_D^T > 0\}\end{equation}
The expression above defines an ellipsoid that is invariant if the system trajectory originating within the ellipsoid remains in that ellipsoid at all times. In terms of Lyapunov function $\mathcal{V}(x)$, this implies {\color{black}\cite{b45,b46}}:
\begin{equation}\label{(28)}\begin{matrix}\dot{\mathcal{V}}(x) \leq 0\ \ \ \ |\ \ \ \ \mathcal{V}(x) \geq 1\end{matrix}\end{equation}
subject to conditions \eqref{(15)}, \eqref{(16a)} and \eqref{(16b)}.

Generally, minimizing the invariant ellipsoid constrains the permissible system trajectories of the perturbed system, making the system ‘less sensitive' to perturbations. Consequently, minimizing the invariant ellipsoid is of interest in the pursuit of robust disturbance rejection control. Soliman and Al-Hinai \cite{b49} reported better results in a robust AGC scheme using the invariant ellipsoid technique than its $H_\infty$ counterpart.

To obtain a Lyapunov function that meets the above requirements, we propose to utilize LMI optimization. Given a decentralized control law of the form $u=K_Dx$, the closed-loop system is guaranteed to be stable if the Lyapunov function in \eqref{(19)} satisfies:
\begin{equation}\label{(29)}\dot{\mathcal{V}} =  x^T\boldsymbol{P}\dot{x} + \dot{x}^T\boldsymbol{P}x \leq 0\end{equation}
Recalling \eqref{(13)}, this becomes:
\begin{equation}\label{(30)}\begin{aligned}\dot{\mathcal{V}}(x) =\ & x^T(A^T\boldsymbol{P} + K^TB^T\boldsymbol{P} + \boldsymbol{P}A + \boldsymbol{P}BK)x \ + \\ & x^T\boldsymbol{P}Ed + d^TE^T\boldsymbol{P}x \ + \\ & x^T\boldsymbol{P}h + h^T\boldsymbol{P}x + x^T\boldsymbol{P}g + g^T\boldsymbol{P}x \leq 0\end{aligned}\end{equation}

\vspace{3.5pt}
\manuallabel{Remark 3}{3}\textit{Remark 3.} To simplify the notation, we dropped the subscript $D$ from all the matrices in \eqref{(29)} and \eqref{(30)}. We will continue to do so throughout this section.
\vspace{5.5pt}

To ensure that the bounds on $\mathcal{V}, d, \Delta A, \Delta E$ are satisfied, we reformulate \eqref{(30)} as follows:
\begin{equation}\label{(31)}\begin{matrix}F_{\dot{\mathcal{V}}} = \dot{\mathcal{V}}(x) \leq 0, & \forall & F_{\mathcal{V}} = -x^T\boldsymbol{P}x \leq -1\end{matrix}\end{equation}
subject to:
\begin{subequations}\begin{align}
&\label{(32a)}F_d = d^T\bar{S}d \leq 1 \\
&\label{(32b)}F_h = h^Th - \alpha^2 x^TH^THx \leq 0 \\
&\label{(32c)}F_g = g^Tg - \beta^2 d^TG^TGd \leq 0\end{align}\end{subequations}
where $F_{\dot{\mathcal{V}}}$ and $F_{\mathcal{V}}$ are Lyapunov conditions and $F_d$, $F_h$ and $F_g$ correspond to inequalities \eqref{(15)}, \eqref{(16a)} and \eqref{(16b)}{\color{black},} respectively. Then, we invoke the S-procedure \cite{b50} to combine these inequalities with {\color{black}respective Lyapunov, disturbance constraint, state and disturbance uncertainty S-procedure} multipliers $\tau_\mathcal{V}, \tau_d, \tau_h, \tau_g$ as:
\begin{equation}\label{(33)}\begin{aligned} & x^T(A^T\boldsymbol{P} + K^TB^T\boldsymbol{P} + \boldsymbol{P}A + \boldsymbol{P}BK)x\ + \\ & x^T\boldsymbol{P}Ed + d^TE^T\boldsymbol{P}x +  x^T\boldsymbol{P}h + h^T\boldsymbol{P}x\ + \\ & x^T\boldsymbol{P}g + g^T\boldsymbol{P}x + \tau_{\mathcal{V}} x^T\boldsymbol{P}x - \tau_dd^T\bar{S}d\ - \\ & \tau_hh^Th + \tau_h\alpha^2x^TH^THx\ - \tau_gg^Tg + \tau_g\beta^2d^TG^TGd \leq 0\end{aligned}\end{equation}
Following Zečević and Šiljak \cite{b51}, this can be rewritten as: 
\begin{equation}\label{(34)}\begin{bmatrix}\mathcal{N} & YH^T & E & 0 & I & I \\ HY & -\tau_h^{-1}\gamma_\alpha I & 0 & 0 & 0 & 0 \\ E^T & 0 & -\tau \bar{S} & G^T & 0 & 0 \\ 0 & 0 & G & -\tau_g^{-1}\gamma_\beta I & 0 & 0 \\ I & 0 & 0 & 0 & -\tau_hI & 0 \\ I & 0 & 0 & 0 & 0 & -\tau_gI\end{bmatrix} \leq 0\end{equation}
where $\tau_\mathcal{V} = \tau_d = \tau$ {\color{black}for a scalar $\tau$}, $\mathcal{N} = YA^T + L^TB^T + AY + BL + \tau Y$, $\gamma_\alpha = \frac{1}{\alpha^2}$, $\gamma_\beta = \frac{1}{\beta^2}$, $L = KY$ and $Y=\boldsymbol{P}^{-1}$.


To produce state feedback control that preserves distributed-averaging consensus control and power sharing, the structure of $\boldsymbol{P}$ and $K$ is set to be:
\begin{equation*}
\boldsymbol{P} = \tiny\begin{matrix} \begin{bNiceArray}{c c:c:c c} 
*\ * & & & & \\ 
*\ * & & & &  \\ 
& *\ * & & & \\ 
& *\ * & & & \\ \hdottedline 
& & \ddots & &   \\ \hdottedline 
& & & *\ * & \\ 
& & & *\ * & \\ 
& & & & *\ * \\ 
& & & & *\ * \end{bNiceArray}
\end{matrix};\ K = \tiny\begin{matrix} \begin{bNiceArray}{c c:c:c c} 
*\ * & & & & \\ 
& *\ * & & & \\ \hdottedline 
& & \ddots &  & \\ \hdottedline 
& & & *\ * & \\ 
& & & & *\ * \end{bNiceArray} \end{matrix}
\end{equation*}
This can be accomplished by choosing the same block structure for matrices $L$ and $Y$. In order to incorporate bounds on parameters $\alpha$ and $\beta$ and ensure that the gain values are realistic, we introduce three additional inequalities:
\begin{equation*}\begin{matrix}
\gamma_\alpha \geq 1, && \gamma_\beta \geq 1, && L^TL \leq \kappa_LI\end{matrix}\end{equation*}
{\color{black}
where $\kappa_L$ is a scalar bound.}

\vspace{5.5pt}
\manuallabel{Remark 4}{4}\textit{Remark 4.} $Y^{-1}$ is bounded by virtue of \eqref{(26)}. Given that $K=LY^{-1}$, it is only necessary to consider bounds on $L$.
\vspace{5.5pt}

As shown in \cite{b51}, the last inequality can be rewritten as:
\begin{equation}\label{(35)}\begin{bmatrix}-\kappa_LI & L^T \\ L & -I_M\end{bmatrix} \leq 0\end{equation}
where $I_M$ is an identity matrix of dimension $M \times M$ and $M$ is the total number of inputs. This produces the LMI optimization problem in its final form {\color{black}with scalar multipliers $c_1,c_2,c_3$} and initial set containment matrix $Y_0$:

\manuallabel{Proposition 1}{1}\textit{Problem 1. Minimize}
\begin{equation}\label{(36)}c_1\gamma_\alpha + c_2\gamma_\beta + c_3\kappa_L\end{equation}

\textit{subject to}
\begin{equation}\label{(37)}Y \geq Y_0 >0\end{equation}
\begin{equation}\label{(38)}\begin{bmatrix}\mathcal{N} & YH^T & E & 0 & I & I \\ HY & -\tau_h^{-1}\gamma_\alpha I & 0 & 0 & 0 & 0 \\ E^T & 0 & -\tau \bar{S} & G^T & 0 & 0 \\ 0 & 0 & G & -\tau_g^{-1}\gamma_\beta I & 0 & 0 \\ I & 0 & 0 & 0 & -\tau_hI & 0 \\ I & 0 & 0 & 0 & 0 & -\tau_gI\end{bmatrix} \leq 0\end{equation}
\begin{equation}\label{(39)}\begin{bmatrix}-\kappa_LI & L^T \\ L & -I_M\end{bmatrix} \leq 0\end{equation}
\begin{equation}\label{(40)}\gamma_\alpha - 1 \geq 0\end{equation}
\begin{equation}\label{(41)}\gamma_\beta - 1 \geq 0\end{equation}
{\color{black}The optimization problem objective is to minimize the sum of three parameters: inverse square frequency consensus connective strength $\gamma_\alpha$, inverse square voltage consensus connective strength $\gamma_\beta$ and state feedback matrix bounding scalar $\kappa_L$. $c_1,c_2,c_3$ are the respective $\gamma_\alpha,\gamma_\beta,\kappa_L$ scalar multipliers. By minimizing $\gamma_\alpha$ and $\gamma_\beta$, we maximize the connective strength $\alpha$ and $\beta$ of the consensus dynamics $\Delta A_D$ and $\Delta E_D$ such that connective stability is guaranteed (typically less than 1). This occurs alongside robustness against disconnections and reconnections of MGs and DERs in NMGs with minimal consensus compromise. Additionally, the optimization problem solves for the inverse Lyapunov function matrix $Y$ (for guaranteed feedback stability) and transformed gain $L=KY$ (for state feedback gain that is constrained by $\kappa_L$).} If a feasible solution is found, the gain matrix can be computed directly as $K=LY^{-1}$.

\vspace{5.5pt}
{\color{black}\manuallabel{Remark 5}{5}\textit{Remark 5.} The proposed control strategy is for the derived DAPI-based NMG model in \eqref{(13)} and bounds \eqref{(15)}, \eqref{(16a)}, \eqref{(16b)} and \eqref{(26)}. Nonetheless, in general, the underlying concepts of connective stability and invariant ellipsoids are applicable to NMG models employing other consensus algorithms.}
\vspace{5.5pt}

\vspace{5.5pt}
{\color{black}\manuallabel{Remark 6}{6}\textit{Remark 6.} For large inverter capacity values, the elements of $\bar{S}$ may become several orders of magnitude smaller than other LMI terms. Thus, to improve numerical conditioning, per-unitized values can be used within the LMI-based problem.}
\vspace{0.5pt}

{\color{black}The flowchart diagram of the proposed scheme is provided in Fig. \ref{Fig. 2}, and a pseudocode algorithm is defined in Algorithm \ref{Algorithm 1}. The ellipsoid $\mathcal{E}_x$ can be minimized for better disturbance rejection. However, for a fair comparison with other control strategies, $\tau$ is bounded at 1. Alternatively, scalar weighting factors can be set to obtain equitable $\alpha$ and $\beta$, with less emphasis on minimizing $K$ (for example, $c_1 = 0.01, c_2 = 10, c_3 = 10$).} The algorithm will compute the connective strength bounds $\alpha$ and $\beta$, along with the state feedback gain matrix $K$. In the following section, we validate the proposed scheme using simulation results.

\begin{figure}[b!]\manuallabel{Fig. 2}{2}\vspace{-8pt}
    \begin{center}
  \includegraphics[width=\columnwidth]{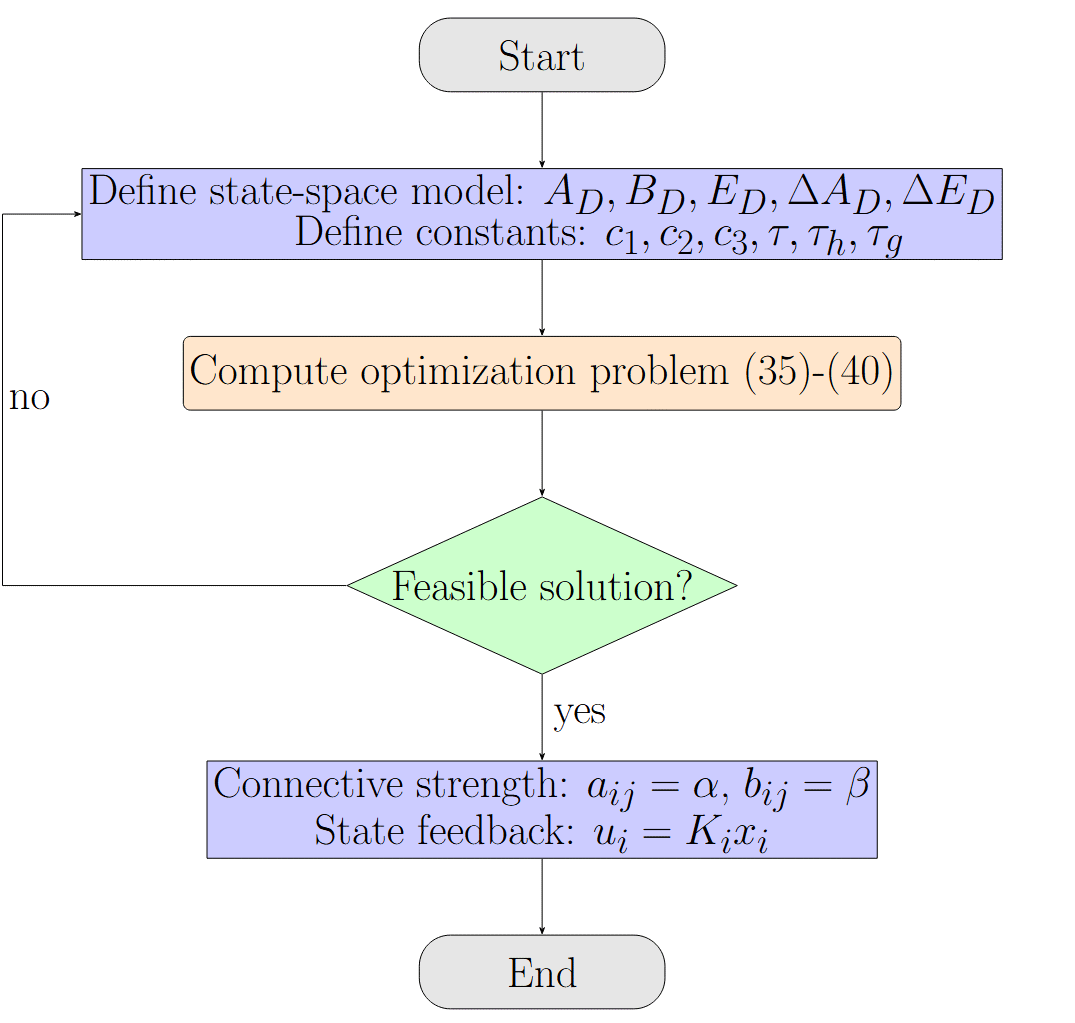}
    \caption{{\color{green}A flowchart of the connective strength and state feedback gain offline calculation process and the online implementation (by defining $a_{ij},b_{ij},K_i$) using the proposed connectively stable and robust control scheme.}}
    \end{center}
\end{figure}


\begin{algorithm}[t!]\manuallabel{Algorithm 1}{1}\small
\caption{\small Connectively Stable Linear Matrix Inequality (LMI)-based Robust Algorithm for Computing the Feedback Gain and Connective Strength Bounds}
\begin{algorithmic}
  \State \textbf{Input:} 
  \State State-space matrices \(A_D \in \mathbb{R}^{4N \times 4N}\), \(B_D \in \mathbb{R}^{4N \times 2N}\), \(E_D \in \mathbb{R}^{4N \times 2N}\), disturbance bound matrix \(\bar{S}_D \in \mathbb{R}^{2N \times 2N}\), uncertainty matrices \(H \in \mathbb{R}^{4N \times 4N}\), \(G \in \mathbb{R}^{4N \times 2N}\) and initial set matrix \(Y_{D,0} \in \mathbb{R}^{4N \times 4N}\).
  \State Define or adjust constants $c_1, c_2, c_3, \tau, \tau_h, \tau_g$.
  
  \State \textbf{Step 1: Declare decision variables}
  \State \quad \(Y_D \geq Y_{D,0}\) (\(Y_D\) is symmetric, i.e., \(Y_D = Y_D^T\) and block diagonal), \(L_D\), \(\gamma_\alpha\), \(\gamma_\beta\), \(\kappa_L\)
  
  \State \textbf{Step 2: Define LMI constraints}
  \State \quad \(C_1-C_5:\; \eqref{(37)}-\eqref{(41)}\)

  \State \textbf{Step 3: Set up the optimization problem}
  \State \quad \(\texttt{Constraints} \gets \{C_1, C_2, C_3, C_4,C_5\}\)
  \State \quad \(\texttt{Objective} \gets \min\ c_1\gamma_\alpha + c_2\gamma_\beta + c_3\kappa_L\)
  
      \State \textbf{Step 4: Solve the LMI optimization problem}
\State \quad \textbf{If solution is feasible:}
    \State \quad \quad \textbf{break}
\State \quad \textbf{else}
    \State \quad \quad goto: \textbf{Input}
\State \quad \textbf{end if}
\State \textbf{Output:}
\State $\boldsymbol{P}_D \gets Y_D^{-1}$, $K_D \gets L_DY_D^{-1}$, $\alpha \gets sqrt(1/\gamma_\alpha)$, $\beta \gets sqrt(1/\gamma_\beta)$
\State \Return $K_D, \boldsymbol{P}_D, \alpha, \beta$
\end{algorithmic}
\end{algorithm}

\manuallabel{Section V}{V}\section{Simulation Results}
To validate the proposed scheme, an NMG consisting of 3 MGs and 5 DERs is modeled and simulated (this system is similar to the one in \cite{b31}). {\color{black}The simulation was performed in MATLAB\textsuperscript{\textregistered}/Simulink\textsuperscript{\textregistered} using the SimPowerSystems toolbox. Each DER is modeled in detail as an example of a distribution-level GFM IBR with BESS, including inner voltage and current control loops, primary level and secondary level control (as shown in Fig. \ref{Fig. 2.1}).} {\color{green}As mentioned in Section} \ref{Section II}{\color{green}, it is assumed that the BESS is pre-charged by a GFL generation source such as a PV system, with enough stored energy for a short transient interval. The generation source and charging dynamics are outside the scope of this work.} The NMG, shown in Fig. \ref{Fig. 2.3}, consists of 3 MGs and balanced three-phase loads. Within the MGs, DERs have communication links with each other, and a {\color{black}communication} link with one of the neighboring MG's DERs. Table \ref{Table 2.1} shows the relevant DER parameters following \cite{b12}. 

{\color{green}As discussed in \cite{b52}, we compute the droop gains as the ratio of the maximum frequency and voltage deviation to the maximum inverter apparent power deviation. Since apparent power capacity is simultaneously allocated for both frequency and voltage droop gain, we set small frequency and voltage deviation terms. Accordingly, we compute the droop gains as follows:

\begin{equation*}
    m_i = \frac{\omega_{max,i}-\omega_{min,i}}{\Delta \bar{s}_i} = \frac{\omega^*+2-(\omega^*-2)}{\Delta \bar{s}_i} = \frac{4}{\Delta \bar{s}_i}\ \frac{rad}{Ws}
\end{equation*}
\begin{equation*}
    n_i = \frac{V_{max,i}-V_{min,i}}{\Delta \bar{s}_i} = \frac{V^*+4-(V^*-4)}{\Delta \bar{s}_i} = \frac{8}{\Delta \bar{s}_i}\ \frac{V}{VAr}
\end{equation*}
$\Delta \bar{s}_i$ is calculated as follows: 
\begin{equation*}
    \Delta \bar{s}_{1} = \Delta \bar{s}_{2} = \Delta \bar{s}_{4} = \Delta \bar{s}_{5} = 2\times20000 = 40000\ VA
\end{equation*}
\begin{equation*}
    \Delta \bar{s}_{3} = 2\times40000 = 80000\ VA
\end{equation*}
Thus, the droop gains are:
\begin{equation*}
    m_1 = m_2 = m_4 = m_5 = \frac{4}{40000} = 1\times 10^{-4}\ \frac{rad}{Ws}
\end{equation*}
\begin{equation*}
    m_3 = \frac{4}{80000} = 0.5\times 10^{-4}\ \frac{rad}{Ws}
\end{equation*}
\begin{equation*}
    n_1 = n_2 = n_4 = n_5 = \frac{8}{40000} = 2\times 10^{-4}\ \frac{V}{VAr}
\end{equation*}
\begin{equation*}
    n_3 = \frac{8}{80000} = 1\times 10^{-4}\ \frac{V}{VAr}
\end{equation*}}

\begin{figure}[t!]\manuallabel{Fig. 2.3}{3}
  \centering

  \begin{minipage}[b]{0.48\textwidth}
    \centering
    \includegraphics[width=\linewidth]{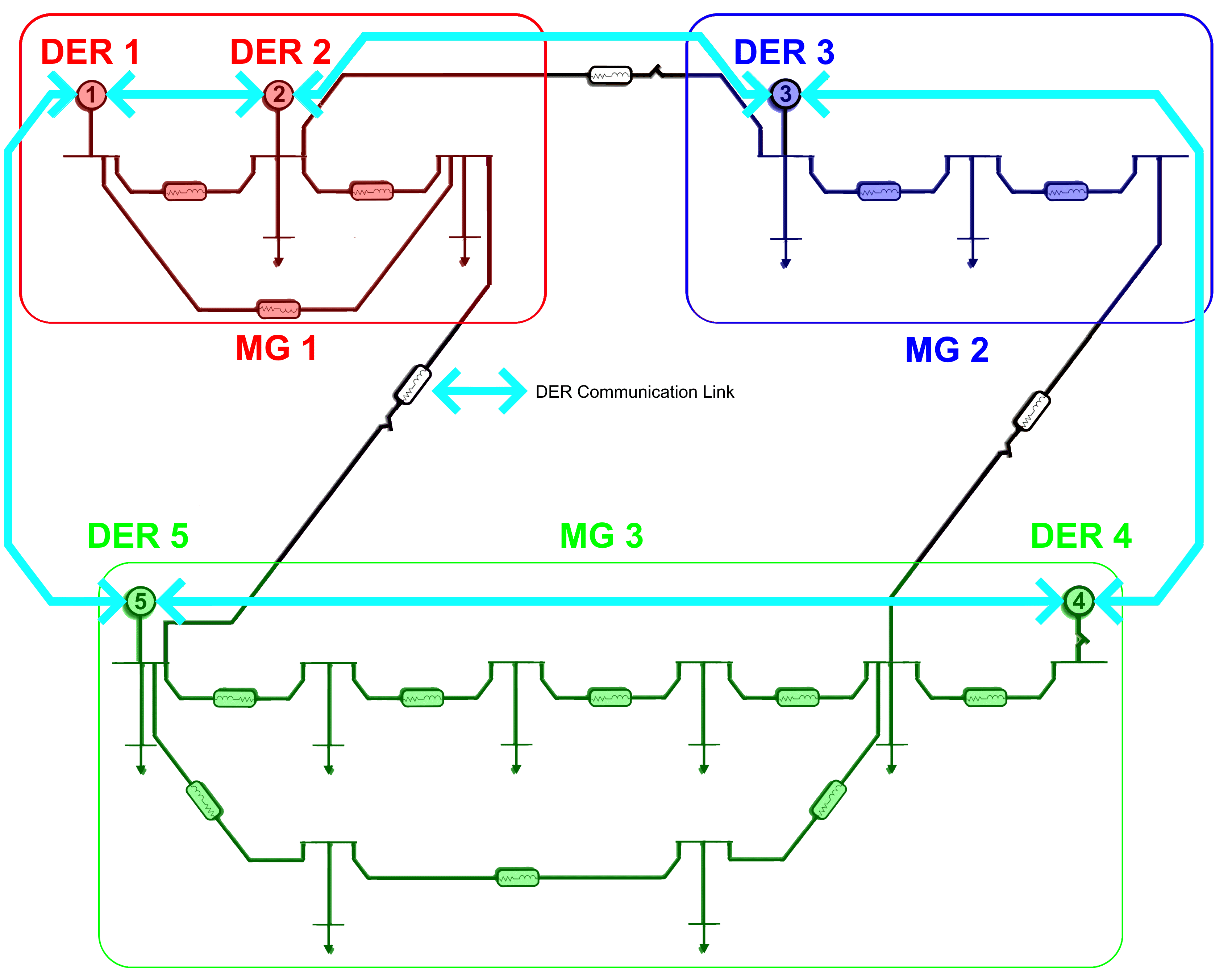}\\
    {\footnotesize {\color{green}(a)}}
  \end{minipage}
  \begin{minipage}[b]{0.48\textwidth}
    \centering
    \includegraphics[width=\linewidth]{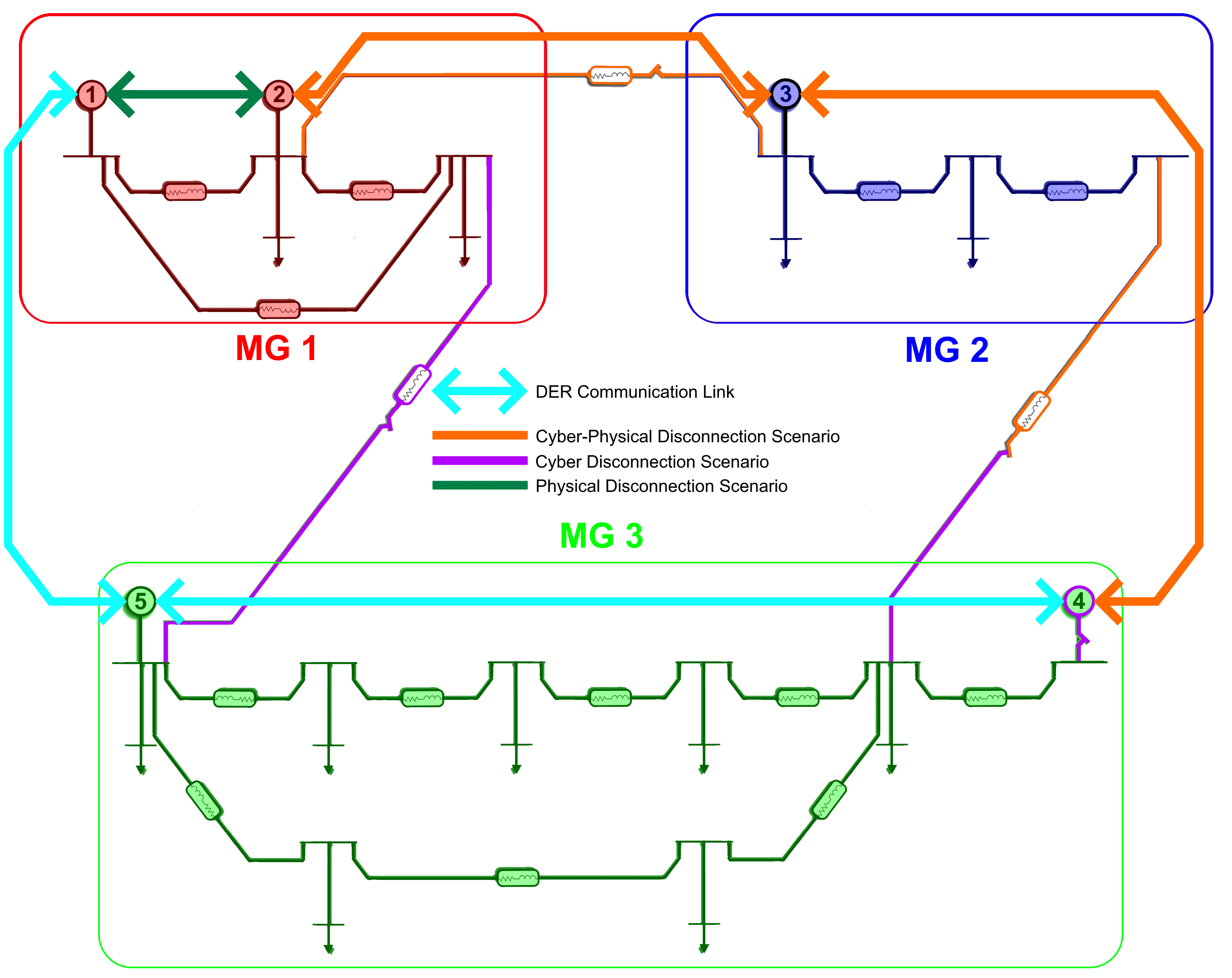}\\[-4pt]
    {\footnotesize {\color{green}(b)}}
  \end{minipage}

  {\color{green}\caption{(a) The Network of Microgrids (NMG) model consisting of 3 Microgrids (MGs) and 5 Distributed Energy Resources (DERs), and (b) the cyber-physical event scenarios. The developed model in \eqref{(13)} represents the aggregated primary and secondary level control of the DERs.}}
  \label{Fig. 1}\vspace{-10pt}
\end{figure}

The DAPI control adapted from \cite{b12} (which we will refer to as the “base {\color{black}DAPI} scheme") is compared with {\color{black}a dissipativity-based scheme, an $H_\infty$-based scheme and the proposed scheme. We refer to Appendices \ref{Appendix B} and \ref{Appendix C} for derivation details. While the proposed scheme's distinguishing advantage is addressing structural perturbations, it is important to note that interconnection matrices are considered to be nominal in the aforementioned baseline approaches, since they are typically applied as robust control schemes for disturbance rejection. We also note that state feedback gains are within $|k_{11}|,|k_{12}|,|k_{21}|,|k_{22}|\leq 1$ range, as excessively large gains can lead to increasingly long settling times. For the $H_\infty$ case, the respective state feedback gains can only produce minimized variations at the cost of long settling times.} 

In Fig. \ref{Fig. 2.4}, the DERs start {\color{black}load pickup} at $t=0$, and include both primary level and secondary level control. To emulate cyber-physical disruptions and test the robustness of the schemes, three {\color{black}further} scenarios have been simulated, in which DERs, MGs and their communication links are subject to cyberattack scenarios. In evaluating the system performance, we applied the following criteria: (1) Confidentiality, (2) Integrity and (3) Availability \cite{b53,b54}.

\begin{table}{\color{black}\manuallabel{Table 2.1}{I}\smaller
\caption{The Network of Microgrids (NMG) Electrical and Control Parameters}
\centering
\begin{tabular}{c|c|c}
\hline
\hline
Parameter & Symbol & Value  \\
\hline
\multicolumn{3}{c}{Distributed Energy Resource (DER) Electrical Parameters}  \\
\hline
{\color{green}Battery rated energy} & {\color{green}$E_{rated,i}$} & {\color{green}68 kWh (DERs 1,2,4,5)} \\
 &  & {\color{green}136 kWh (DER 3)} \\
{\color{green}Initial state of charge (SOC)} & {\color{green}$SoC$} & {\color{green}50$\%$} \\
Inverter Switching frequency & $f_{sw}$ & 10 kHz\\
Nominal frequency & $\omega^*/2\pi$ & 60 Hz\\
Nominal AC voltage & $V^*$ & 169.7 V\\
DC voltage & $V_{dc}$ & 339.4 V\\
Inverter-side filter resistance & $R_{f,inv}$ & 0.1 $\Omega$ \\
Inverter-side filter inductance & $L_{f,inv}$ & 1.8 mH \\
Output-side filter resistance & $R_{f,pcc}$ & 0.1 $\Omega$ \\
Output-side filter inductance & $L_{f,pcc}$ & 1.8 mH \\
Filter Capacitance & $C_f$ & 25 $\mu F$\\
\hline
\multicolumn{3}{c}{Distributed Energy Resource (DER) Control Parameters}  \\
\hline
Setpoint active power & $P^*_i$ & 10 kW (DERs 1,2,4,5) \\
 &  & 20 kW (DER 3) \\
Setpoint reactive power & $Q^*_i$ & 0.5 kVAr (DERs 1,2,4,5) \\
  &   & 1 kVAr (DER 3) \\
Frequency consensus link gain & $a_{ij}$ & 1 \\
Voltage consensus link gain & $b_{ij}$ & 1 \\
Inverse integral frequency gain  & $k_i$ & 1 \\
Integral voltage gain  & $\kappa_i$ & 1 \\
\hline
\multicolumn{3}{c}{Network of Microgrids (NMG) System Parameters}  \\
\hline
Line impedance A & $Z_A$ & R=0.1 $\Omega$, L = 1.8 mH \\
Line impedance B & $Z_B$ & R=0.4 $\Omega$, L = 1.8 mH \\
Line impedance C & $Z_C$ & R=0.35 $\Omega$, L = 1.9 mH \\
MG 1 load & $S_1$ & 13.5 kW + j5.6 kVAr \\
MG 2 load & $S_2$ & 15.9 kW + j5.4 kVAr \\
MG 3 load & $S_3$ & 14.1 kW + j5.2 kVAr \\
\hline
\hline
\end{tabular}
\vspace{-8pt}
}\end{table}\normalsize

\begin{figure*}[h!]\centering\manuallabel{Fig. 2.4}{4}
  \includegraphics[width=0.95\textwidth]{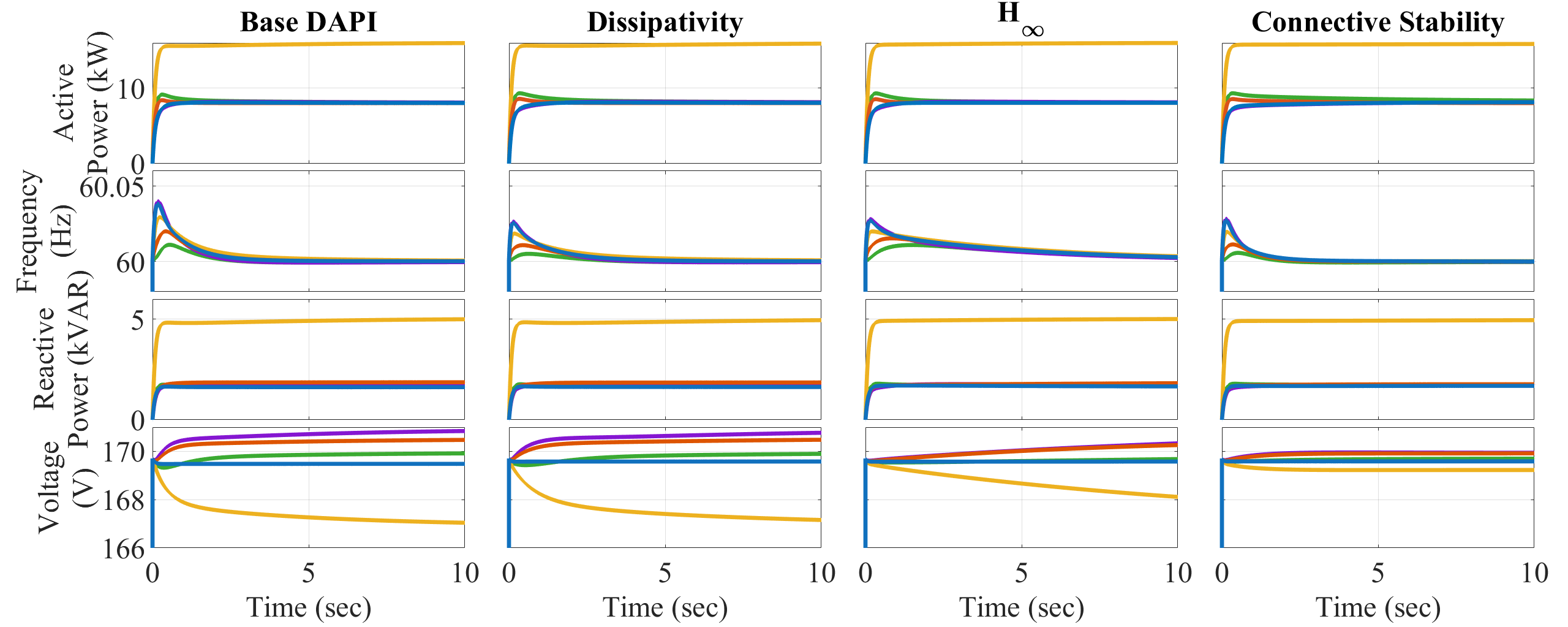}
  \caption{{\color{black}Comparison of four different methods for the five Distributed Energy Resources (DERs) during load pickup. From top to bottom: Distributed-Averaging Proportional-Integral (DAPI)-based DER active power, frequency, reactive power and voltage from 0 to 10 seconds for \legline{der1}{DER 1}, \legline{der2}{DER 2}, \legline{der3}{DER 3},
\legline{der4}{DER 4}, and \legline{der5}{DER 5}.}}
  \vspace{-4pt}
\end{figure*}

\subsection{Cyber-Physical (Confidentiality) Islanding: Disconnecting Resources}
In our simulations, we considered a scenario akin to a confidentiality cyberattack, in which an adversary gained full access to the NMG control network. After 10 seconds, both the physical and communication links of MG 2 were disconnected from the rest of the NMG, causing an imbalance with small frequency and voltage deviations. The results are displayed in Fig. \ref{Fig. 2.5}{\color{black}. Though the results show somewhat slower power sharing, there is a slight improvement in frequency and voltage regulation with the proposed robust and connectively stable method.}

\begin{figure*}\centering\manuallabel{Fig. 2.5}{5}
  \includegraphics[width=0.95\textwidth]{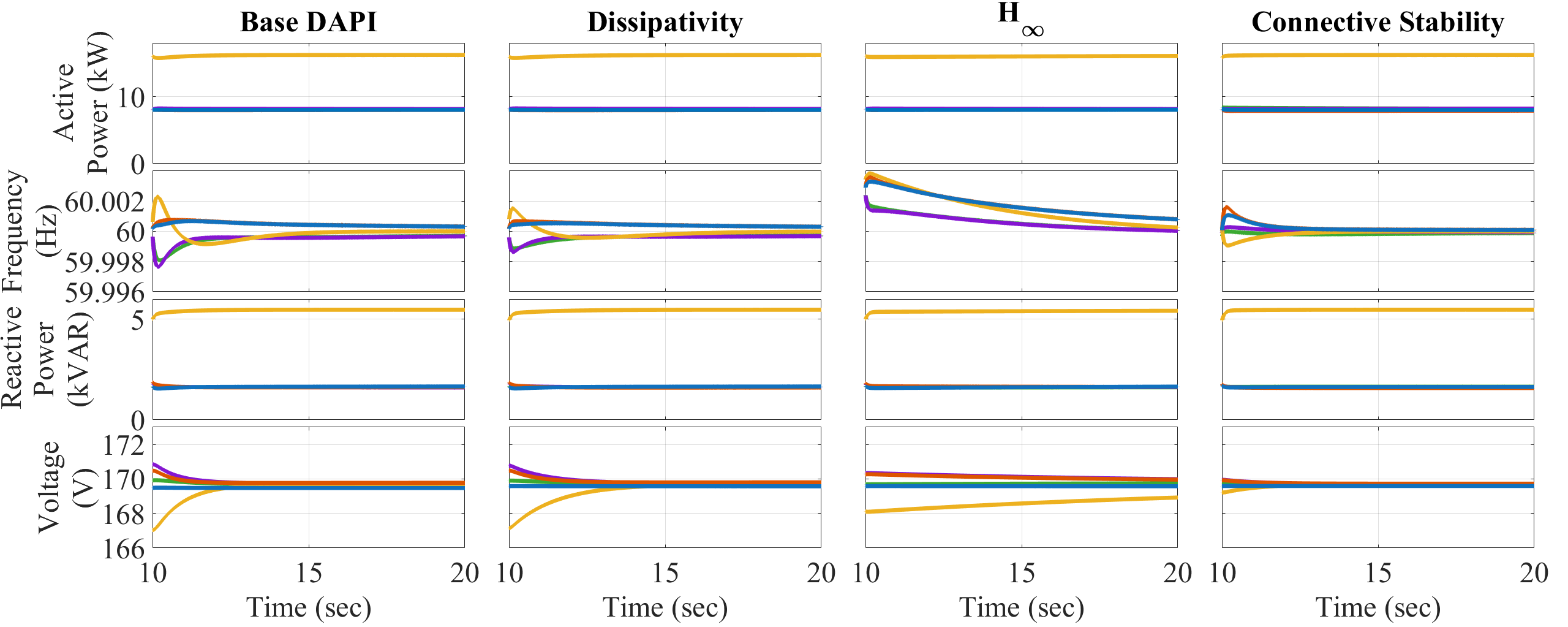}\vspace{-6pt}
  \caption{{\color{black}Comparison of four different methods for the five Distributed Energy Resources (DERs) during cyber-physical disconnection. From top to bottom: Distributed-Averaging Proportional-Integral (DAPI)-based DER active power, frequency, reactive power and voltage from 10 to 20 seconds for \legline{der1}{DER 1}, \legline{der2}{DER 2}, \legline{der3}{DER 3}, \legline{der4}{DER 4}, and \legline{der5}{DER 5}.}}
  \vspace{-4pt}
\end{figure*}

\subsection{Physical (Integrity) Islanding: False Data Injection}
In the next case, at 10 seconds, MG 3 along with DER 4 were physically islanded while the communication links remained unchanged. From an integrity perspective, this scenario resembles an FDI attack, since communication links are transmitting information that is not equivalent to the physical links among DERs. We modeled this scenario by setting: 
\begin{equation*}(a_{ij},b_{ij}) = \begin{cases} 
      \left(\alpha,\beta\right), & \textit{if no FDI occurs}, \\
      \left(\begin{matrix}\alpha + \Delta a_{ij}^{FDI}, \\ \beta + \Delta b_{ij}^{FDI}\end{matrix}\right), & \textit{otherwise}.
\end{cases}\end{equation*}
where {\color{black}$\Delta a_{ij}^{FDI} = \alpha$ and $\Delta b_{ij}^{FDI} = \beta$}. Such a process results in offsets in the consensus control, and consequently, a steady-state error of the frequency and voltage. Additionally, it results in a case setting that is outside the certified bounds of the robust control schemes, for the sake of further robustness validation. Fig. \ref{Fig. 2.6} indicates that the {\color{black}proposed robust and connectively stable method produces the lowest frequency and voltage over-/undershoots and steady-state error}. 

\begin{figure*}\centering\manuallabel{Fig. 2.6}{6}
  \includegraphics[width=0.95\textwidth]{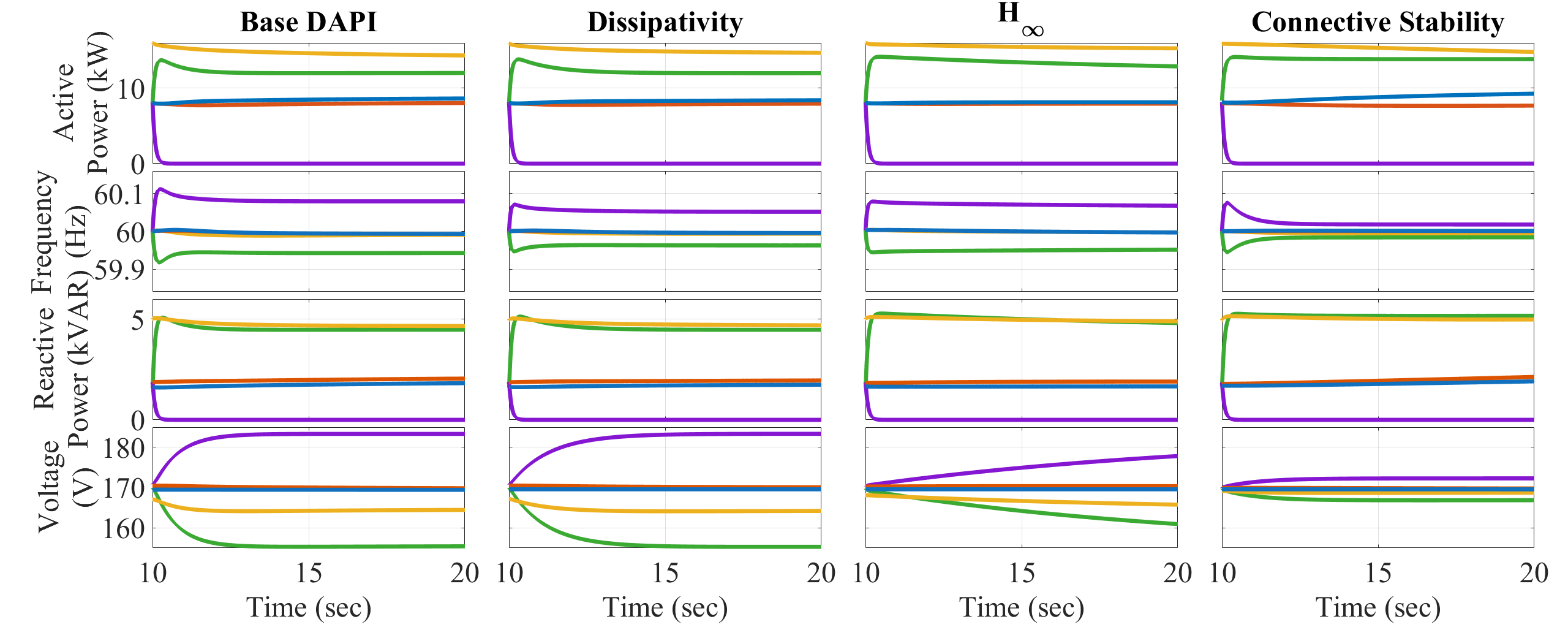}\vspace{-6pt}
  \caption{{\color{black}Comparison of four different methods for the five Distributed Energy Resources (DERs) during physical disconnection. From top to bottom: Distributed-Averaging Proportional-Integral (DAPI)-based DER active power, frequency, reactive power and voltage from 10 to 20 seconds for \legline{der1}{DER 1}, \legline{der2}{DER 2}, \legline{der3}{DER 3},
\legline{der4}{DER 4}, and \legline{der5}{DER 5}.}}
  \vspace{-4pt}
\end{figure*}

\subsection{Cyber (Availability) Islanding: Denial of Service}
As a final step in the simulation, we assume an unintentional islanding event for communication links occurs at 10 seconds. From a cybersecurity perspective, this scenario resembles a Denial of Service (DoS) attack, where communication links that transmit neighboring DAPI term values $\Omega_j{\color{black},Q_j/Q_j^*}$ are essentially shut down, and the communication link between DERs 1 and 2 is disabled. Such a situation can be represented by setting:

\begin{equation*}(a_{ij},b_{ij}) = \begin{cases} 
      \left(\alpha,\beta\right), & \textit{if no DoS occurs}, \\
      \mathbf{0}, & \textit{otherwise}.
\end{cases}\end{equation*}

\begin{figure*}\centering\manuallabel{Fig. 2.7}{7}
  \includegraphics[width=0.95\textwidth]{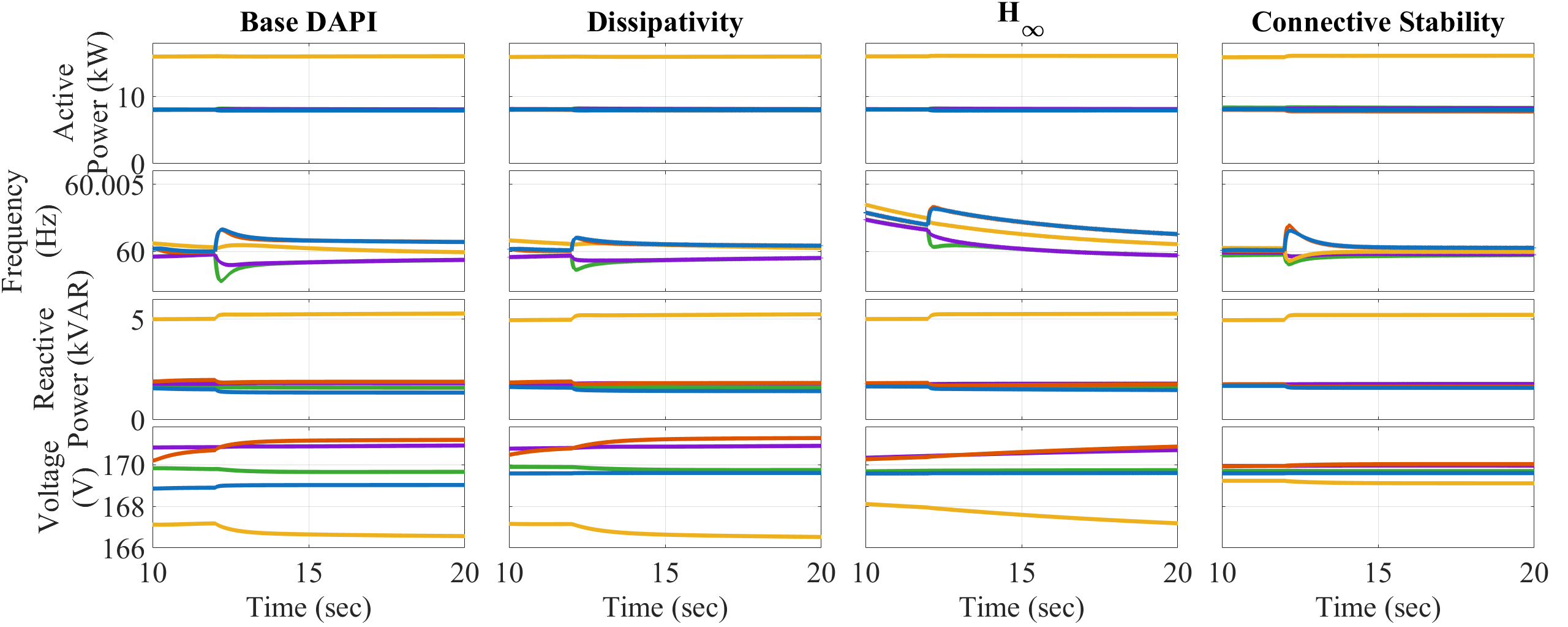}\vspace{-6pt}
  \caption{{\color{black}Comparison of four different methods for the five Distributed Energy Resources (DERs) during cyber disconnection. From top to bottom: Distributed-Averaging Proportional-Integral (DAPI)-based DER active power, frequency, reactive power and voltage from 10 to 20 seconds for \legline{der1}{DER 1}, \legline{der2}{DER 2}, \legline{der3}{DER 3},
\legline{der4}{DER 4}, and \legline{der5}{DER 5}.}}
  \vspace{-4pt}
\end{figure*}

After 12 seconds, MG 1 is islanded. {\color{black}Fig. \ref{Fig. 2.7} indicates that there is a noticeable reduction in steady-state deviation for both the frequency and the voltage when the proposed robust and connectively stable method is applied.} It should also be noted that the power output settling time and sharing deteriorate in all cases, since the loss of the communication link between DERs 1 and 2 leads to inaccurate consensus dynamics.


\subsection{Analysis of the Results}
To evaluate the improvements achieved using the proposed {\color{black}robust and connectively stable} method, we adopted metrics similar to those in \cite{b55,b56} and \cite{b57}. These authors introduced the following loss metric, which reflects {\color{black}frequency $\omega_{Ro}$ and voltage $V_{Ro}$} robustness with respect to disruptions:
\begin{equation}\label{(42)}{\color{black}\begin{split}&\omega_{Ro} =  \frac{1}{N}\sum_{i=1}^N\sup\left(\left|\frac{\omega ^*- \omega_i}{\omega_i}\right|\right), \\ &V_{Ro} =  \frac{1}{N}\sum_{i=1}^N\sup\left(\left|\frac{V ^*- V_i}{V_i}\right|\right)\end{split}}\end{equation}
In \eqref{(42)}, {\color{black}$\omega_i,V_i$ and $\omega^*,V^*$ are the frequency and voltage actual and setpoint values,} respectively. Such a measure is useful because smaller frequency and voltage deviations mean smaller losses, and by extension more robustness with respect to disturbances. 

A {\color{black} resilience performance measure for frequency $\omega_{Re}$ and voltage $V_{Re}$} in which the supremum is replaced by the integral of the deviation over time {\color{black}period} $T$ is another indicator for the effectiveness of a robust control scheme. Such a measure can be defined as:
\begin{equation}\label{(43)}{\color{black}\begin{split}&\omega_{Re} = \frac{1}{N}\sum_{i=1}^N \frac{1}{T}\int_0^T\left|\frac{\omega ^*- \omega_i}{\omega_i}\right| dt, \\ &V_{Re} =  \frac{1}{N}\sum_{i=1}^N\frac{1}{T}\int_0^T\left|\frac{V ^*- V_i}{V_i}\right| dt\end{split}}\end{equation}

\begin{table*}[!t]\color{black}\manuallabel{Table 2.2}{II}\small
\centering
\caption{Frequency $\omega_{Ro}$ and Voltage $V_{Ro}$ Robustness (Supremum) Loss Metric \eqref{(42)} of base Distributed-Averaging Proportional-Integral (DAPI), Dissipativity-based, $H_\infty$-based and the proposed connective stability-based control strategies}
\begin{tabular}{ c c c c c c c c c}
\hline
\hline
 & \multicolumn{4}{c}{\textbf{Peak Frequency Deviation}} & \multicolumn{4}{c}{\textbf{Peak Voltage Deviation}} \\
  & \multicolumn{4}{c}{\textbf{Metric $\omega_{Ro}$ ($\times 10^{-3}$)}} & \multicolumn{4}{c}{\textbf{Metric $V_{Ro}$ ($\times 10^{-3}$)}} \\
\hline
\textbf{Case} & \textbf{Base} & \textbf{Dissipativity} & $H_\infty$ & \textbf{Connective} & \textbf{Base} & \textbf{Dissipativity} & $H_\infty$ & \textbf{Connective}\\
& \textbf{DAPI} & & & \textbf{Stability} & \textbf{DAPI} & & & \textbf{Stability}\\
\hline
\hline
\textbf{Load Pickup} & 0.462 & 0.289 & 0.338 & 0.308 & 6.142 & 5.685 & 3.634 & 1.420 \\
\hline
\textbf{Cyber-physical Disconnection} & 0.027 & 0.018 & 0.052 & 0.014 & 5.954 & 5.593 &  3.517 & 1.276 \\
\hline
\textbf{Physical Disconnection} & 0.729 & 0.464 & 0.487 & 0.463 & 41.44 & 41.38 & 25.74 & 8.045 \\
\hline
\textbf{Cyber Disconnection} & 0.024 & 0.017 & 0.049 & 0.019 & 8.080 & 7.435  & 5.734 & 1.549 \\
\hline
\textbf{Average} & \textbf{0.311} & \textbf{0.197} & \textbf{0.232} & \textbf{0.201} & \textbf{15.40} & \textbf{15.02} & \textbf{9.656} & \textbf{3.073} \\
\hline
\hline
\end{tabular}
\end{table*}

\begin{table*}[!h]\color{black}\manuallabel{Table 2.3}{III}\small
\centering
\caption{Frequency $\omega_{Re}$ and Voltage $V_{Re}$ Resilience (Cumulative) Loss Metric \eqref{(43)} of base Distributed-Averaging Proportional-Integral (DAPI), Dissipativity-based, $H_\infty$-based and the proposed connective stability-based control strategies}
\begin{tabular}{ c c c c c c c c c}
\hline
\hline
 & \multicolumn{4}{c}{\textbf{Cumulative Frequency Deviation}} & \multicolumn{4}{c}{\textbf{Cumulative Voltage Deviation}} \\
  & \multicolumn{4}{c}{\textbf{Metric $\omega_{Re}$ ($\times 10^{-3}$)}} & \multicolumn{4}{c}{\textbf{Metric $V_{Re}$ ($\times 10^{-3}$)}} \\
\hline
\textbf{Case} & \textbf{Base} & \textbf{Dissipativity} & $H_\infty$ & \textbf{Connective} & \textbf{Base} & \textbf{Dissipativity} & $H_\infty$ & \textbf{Connective}\\
& \textbf{DAPI} & & & \textbf{Stability} & \textbf{DAPI} & & & \textbf{Stability}\\
\hline
\hline
\textbf{Load Pickup} & 0.056 & 0.047 & 0.137 & 0.029 & 5.082 & 4.664 & 2.181 & 1.197 \\
\hline
\textbf{Cyber-physical Disconnection} & 0.008 & 0.006 & 0.021 & 0.003 & 0.919 & 1.040 & 2.492 & 0.317 \\
\hline
\textbf{Physical Disconnection} & 0.530 & 0.343 & 0.423 & 0.160 & 37.92 & 36.68 & 16.45 & 7.277 \\
\hline
\textbf{Cyber Disconnection} & 0.010 & 0.008 & 0.023 & 0.005 & 7.349 & 6.731 & 4.653 & 1.450 \\
\hline
\textbf{Average} & \textbf{0.151} & \textbf{0.101} & \textbf{0.151} & \textbf{0.049} & \textbf{12.82} & \textbf{12.28} & \textbf{6.440} & \textbf{2.560}\\
\hline
\hline
\end{tabular}
\end{table*}

{\color{black}We outline the robustness and resilience loss metrics for each control strategy across all scenarios in Table \ref{Table 2.2} and Table \ref{Table 2.3}. Table \ref{Table 2.2} illustrates the peak frequency and voltage deviation metric, a measure relating to over-/undershoot. Table \ref{Table 2.3} illustrates the cumulative frequency and voltage deviation metric, a measure relating to settling time and steady-state error. An average of the loss metrics for all scenarios is provided for a holistic comparison of each control strategy. According to Table \ref{Table 2.2} and Table \ref{Table 2.3}, applying the proposed robust and connectively stable method improves robustness and resilience on average in all considered scenarios. This is especially true when the connective strength of communication links is weaker.

On the other hand, power sharing isn't as efficient as in the cases of the other approaches. This is partly due to the LMI optimization producing the same values of parameters $\alpha$ and $\beta$ for all DERs. Given that connective strength is directly proportional to the speed of power sharing \cite{b12}, it is not surprising that the proposed approach was less effective at suppressing power sharing transients than the other strategies. This can lead to increasingly uneven energy (dis)charge across the BESS-based DERs, which will be addressed in our future work. However, it is important to recognize that power sharing itself across DERs is not compromised. In essence, the DERs' energy expenditure differs between the baseline DAPI scheme and the robust DAPI schemes. This is because with DAPI control, there is a tradeoff involving frequency and voltage transients (over-/undershoot and settling time) and speed of power sharing \cite{b12}. Lastly, reactive power is relatively low in distribution loads, and thus, voltage regulation can take precedence over reactive power sharing. Nonetheless, the proposed scheme is far more effective at minimizing frequency and voltage deviations following cyber-physical perturbation events compared to base DAPI, dissipativity-based and $H_\infty$-based approaches. This is due to the added degree of freedom of tuning the connective strength, making the proposed control scheme less sensitive to perturbations.

Another limitation is that the scheme needs to be recomputed whenever new cyber-physical links are added in the NMG. However, from a practical perspective, the scheme does not require many changes in established control dynamics beyond changing scalar multipliers for specific variables of a droop-controlled inverter with DAPI control. It can be computed once offline to calculate the state feedback gains and connective strength values. Such values translate to scalar multipliers, which provide ease of practical implementation. We do assume that adding new cyber-physical links in the NMG would require recomputation. Since the optimization problem can be solved once offline, implementation can be straightforward, especially as inverter hierarchical control schemes are software-defined in practice. Lastly, Zečević and Šiljak \cite{b51} introduced computationally efficient procedures for decomposing large dynamic systems in LMI algorithms using numerical decomposition techniques. 

We also note that during FDI or DoS events, the proposed method does not require additional control logic or alterations besides the defined scalar parameters (state feedback gains and connective strengths). Computing those parameters can also be done offline, providing a passive approach that is less sensitive to cyberattacks, contrary to existing methods requiring specific modeling or prior data \cite{b54}. Further work might be needed for validation during severe disturbances.

Lastly, further simulations, such as plug-and-play operation, were not tested. Per Simpson-Porco \textit{et al.}'s \cite{b12} work, plug-and-play means DER reconnection following a synchronization process. We did not consider the simulations due to the additional dynamics and control logic needed for synchronization. Our future work regarding distributed synchronization processes will address robust synchronization and connection.}

\section{Conclusions and Future Work}\raisebox{5ex}[0pt][0pt]{\manuallabel{Section VI}{VI}}
A robust distributed secondary level control scheme has been designed and implemented for droop-controlled inverters with DAPI frequency and voltage control. The control scheme utilizes the concept of connective stability to compute the state feedback gain (for the frequency and voltage dynamics) and the connective strength bounds to ensure robustness to operational and structural perturbations. Additionally, disturbance rejection is achieved using the invariant ellipsoid technique by {\color{black}defining an} ellipsoid bound. Simulation of an NMG model consisting of 3 MGs and 5 DERs in MATLAB\textsuperscript{\textregistered}/Simulink\textsuperscript{\textregistered} demonstrates the potential benefits of the proposed scheme{\color{black}. The proposed scheme results in minimized frequency and voltage deviations following cyber-physical perturbation events, when compared to base DAPI, dissipativity-based and $H_\infty$-based approaches. The minimization of robustness and resilience loss metrics validates the performance of the proposed connectively stable and robust scheme.}

Our results suggest that slower power sharing may be a necessary trade-off in the pursuit of achieving robust control that is capable of handling worst-case scenarios of disconnections. From a practical point of view, this trade-off improves primary level and secondary level robustness at the cost of tertiary level energy management. We should note in this context that there is a limit to the degree of robustness that can be achieved by the proposed strategy, as there are capacity limitations that must be considered for small-scale DERs in NMGs. Since such limitations are typically considered at the tertiary level, our future work will focus on developing a hierarchical control framework for regulating the injected energy resulting from primary level and secondary level control. This will allow us to quantify the dynamic reserves of each MG on the primary, secondary and tertiary levels. {\color{black}Further future work involves performance-complexity analysis with other control techniques, plug-and-play operation, controller hardware validation beyond droop control at the primary level and energy sharing control.}

\manuallabel{Section VII}{VII}\section{Acknowledgment}
This work was partially funded by the U.S. National Science Foundation under Grant ECCS-2236843 and the Santa Clara University School of Engineering Dean's Excellence Packard Fellowship.

\section*{Appendices}\renewcommand{\theequation}{A.\arabic{equation}}
\setcounter{equation}{0}

\manuallabel{Appendix A}{A}\section*{Appendix A: Proof of Inequality \eqref{(21)}}
Equation \eqref{(21)} in its expanded form can be derived from \eqref{(13)} and then combined (using the S-procedure \cite{b50}) with \eqref{(15)}, \eqref{(16a)} and \eqref{(16b)} for a Lyapunov function \eqref{(19)} such that $\dot{\mathcal{V}}(x) \leq 0, \mathcal{V}(x) \geq 1$. This is the same process used in Section \ref{Section IV-B} to derive \eqref{(33)}:
\begin{equation*}\begin{aligned}& x^T(A^T\boldsymbol{P} + K^TB^T\boldsymbol{P} + \boldsymbol{P}A + \boldsymbol{P}BK)x\ + \\ & x^T\boldsymbol{P}Ed + d^TE^T\boldsymbol{P}x +  x^T\boldsymbol{P}h + h^T\boldsymbol{P}x\ + \\ & x^T\boldsymbol{P}g + g^T\boldsymbol{P}x + \tau_{\mathcal{V}} x^T\boldsymbol{P}x - \tau_dd^T\bar{S}d\ - \\ & {\color{black}\tau_h}h^Th + {\color{black}\tau_h}\alpha^2x^TH^THx\ - \\ & {\color{black}\tau_g}g^Tg + {\color{black}\tau_g}\beta^2d^TG^TGd \leq 0\end{aligned}\end{equation*}
Reorganizing the equation above {\color{black}for $\tau_\mathcal{V} = \tau_d = \tau$}, we obtain:{\color{black}
\begin{equation}\label{(A.1)}\begin{aligned}& -x^T(-A^T\boldsymbol{P} - K^TB^T\boldsymbol{P} - \boldsymbol{P}A - \boldsymbol{P}BK - \tau_{\mathcal{V}} \boldsymbol{P}\ - \\ & \tau_h\alpha^2H^TH)x + x^T(\Delta A^TP + P\Delta A - \tau_h\Delta A^T\Delta A)x\ + \\ & d^T(E^T\boldsymbol{P})x + x^T(PE)d\ + \\ & d^T(\Delta E^T\boldsymbol{P})x\ + x^T(P\Delta E)d\ + \\ & d^T(-\tau\bar{S} + \tau_g\beta^2G^TG)d + d^T(-\tau_g\Delta E^T\Delta E)d \leq 0\end{aligned}\end{equation}
For $\mathbf{Q}=-\tau\bar{S} + \tau_g\beta^2G^TG$, $\mathbf{S} = -A^T\boldsymbol{P} - K^TB^T\boldsymbol{P} - \boldsymbol{P}A - \boldsymbol{P}BK - \tau_{\mathcal{V}} \boldsymbol{P} - \tau_h\alpha^2H^TH, \mathbf{R} = E^T\boldsymbol{P}, \Delta\mathbf{Q} = -\tau_g\Delta E^T\Delta E, \Delta \mathbf{S} = \Delta A^TP + P\Delta A - \tau_h\Delta A^T\Delta A$, $\Delta\mathbf{R} = \Delta E^T\boldsymbol{P}$,} we obtain \eqref{(21)}. 
\qedsymbol

\manuallabel{Appendix B}{B}\section*{Appendix B: Dissipativity-based Control Problem}
Per Boyd \textit{et al.} \cite{b50}, for $\Delta A_D, \Delta E_D$ being absorbed into $A_D, E_D$, output $z = C_Dx + E_{z,D}d$ with output matrix $C_D = diag\{C_{1},C_{2}\dots,C_N\}$, element $c_D$, $E_{z,D} = \epsilon I_D$ for a sufficiently small scalar $\epsilon$ and:
\begin{equation*}
    C_i = \begin{matrix}
        \begin{bmatrix}
        c_D & 0 & 0 & 0 \\ 0 & 0 & c_D & 0
    \end{bmatrix}, & i \in 1,2,\dots,N\end{matrix}
\end{equation*}
the NMG model in \eqref{(13)} is said to be $\eta$ dissipative if:
\begin{equation}\label{(A.3)}
    \begin{matrix}
        \int_0^T (d^TC_Dx + d^TE_{z,D}d - \eta d^Td)\ dt \geq 0, & 0 \leq t \leq T
    \end{matrix}
\end{equation}
where $\eta$ is the dissipation scalar factor. Suppose there exists Lyapunov function \eqref{(19)} such that \cite{b50}:
\begin{equation}\label{(A.4)}
    \dot{\mathcal{V}} \leq  2d^TC_Dx + 2d^TE_{z,D}d - 2\eta d^Td
\end{equation}
Dropping subscript D and following Section \ref{Section IV-B}, a capacity-constrained dissipativity-based problem can be defined as follows for $\alpha = \beta = 1$ and initial set containment matrix $Y_0$:
\begin{equation}\label{(A.5)}\min_{\eta,Y,L,\kappa_L}\ -c_1\eta + c_2\kappa_L\end{equation}

\textit{subject to}
\begin{equation}\label{(A.6)}Y \geq Y_0 >0\end{equation}
\begin{equation}\small\label{(A.7)}\begin{bmatrix}\mathcal{N}  & E-YC^T  \\ E^T-CY & 2\eta I - (E_z^T + E_z) -\tau \bar{S}\end{bmatrix} \leq 0\end{equation}
\begin{equation}\label{(A.8)}\begin{bmatrix}-\kappa_LI & L^T \\ L & -I_M\end{bmatrix} \leq 0\end{equation}
where $\mathcal{N} = YA^T + L^TB^T + AY + BL + \tau Y$, $L = KY$, $Y=\boldsymbol{P}^{-1}$.

\manuallabel{Appendix C}{C}\section*{Appendix C: $H_\infty$-Based Control Problem}
The $H_\infty$ norm of the NMG model in \eqref{(13)} can be defined as $||C_D(sI-(A_D+B_DK_D))^{-1}E_D||_\infty$ for $\Delta A_D, \Delta E_D$ being absorbed into $A_D, E_D$, output $z=C_Dx$ with output matrix $C_D = diag\{C_{1},C_{2}\dots,C_N\}$, element $c_D$ and:
\begin{equation*}
    C_i = \begin{matrix}
        \begin{bmatrix}
        c_D & 0 & 0 & 0 \\ 0 & 0 & c_D & 0
    \end{bmatrix}, & i \in 1,2,\dots,N\end{matrix}
\end{equation*}
Alternatively, per Boyd \textit{et al.} \cite{b50}, the $H_\infty$ norm is equivalent to the $L_2$ gain of the same model. The $L_2$ gain is defined and bounded by scalar $\gamma$ as follows \cite{b50}: 
\begin{equation}\label{(A.9)}
    \begin{matrix}
        \sup_{||d||_2\neq 0} \frac{||C_Dx||_2}{||d||_2} \leq \gamma, & \gamma \geq 0
    \end{matrix}
\end{equation}
Suppose there exists Lyapunov function \eqref{(19)} such that \cite{b50}:
\begin{equation}\label{(A.10)}
    \dot{\mathcal{V}} \leq - x^TC_D^TC_Dx + \gamma^2d^Td
\end{equation}
Dropping subscript D and following Section \ref{Section IV-B}, the capacity-constrained $H_\infty$-based problem can be defined as follows for $\alpha = \beta = 1$, initial set containment matrix $Y_0$ and $\rho$ defined as $\gamma^2$:
\begin{equation}\label{(A.11)}\min_{\rho,Y,L,\kappa_L}\ c_1\rho + c_2\kappa_L\end{equation}

\textit{subject to}
\begin{equation}\label{(A.12)}Y \geq Y_0 >0\end{equation}
\begin{equation}\small\label{(A.13)}\begin{bmatrix}\mathcal{N} & E & YC^T \\ E^T & -\tau \bar{S} - \rho I & 0 \\ CY & 0 & -I\end{bmatrix} \leq 0\end{equation}
\begin{equation}\label{(A.14)}\begin{bmatrix}-\kappa_LI & L^T \\ L & -I_M\end{bmatrix} \leq 0\end{equation}
where $\mathcal{N} = YA^T + L^TB^T + AY + BL + \tau Y$, $L = KY$ and $Y=\boldsymbol{P}^{-1}$.


\begin{thebibliography}{00}
\bibitem{b1} B. Chen, J. Wang, X. Lu, C. Chen and S. Zhao, “Networked Microgrids for Grid Resilience, Robustness, and Efficiency: A Review," IEEE Transactions on Smart Grid, vol. 12, no. 1, pp. 18-32, Jan. 2021.
\bibitem{b2} “IEEE Guide for Control and Automation Installations Applied to the Electric Power Infrastructure," IEEE Std 2030.4-2023, pp. 1-39, 22 Nov. 2023.
\bibitem{b3} M. Ferrari, B. Ollis, M. Starke, A. Massol-Deya, “Why the Next Microgrids Will Be Well Connected," IEEE Spectrum Magazine, vol. 60, no. 10, pp. 36-43, Oct. 2023.
\bibitem{b4} G. Ahmad, A. Hassan, A. Islam, M. Shafiullah, M. A. Abido and M. Al-Dhaifallah, “Distributed Control Strategies for Microgrids: A Critical Review of Technologies and Challenges," IEEE Access, vol. 13, pp. 60702-60719, 2025.
\bibitem{b5} M. N. Alam, S. Chakrabarti and A. Ghosh, “Networked Microgrids: State-of-the-Art and Future Perspectives," IEEE Transactions on Industrial Informatics, vol. 15, no. 3, pp. 1238-1250, March 2019.
\bibitem{b6} Q. Zhou, M. Shahidehpour, A. Paaso, S. Bahramirad, A. Alabdulwahab and A. Abusorrah, “Distributed Control and Communication Strategies in Networked Microgrids," IEEE Communications Surveys \& Tutorials, vol. 22, no. 4, pp. 2586-2633, Fourth Quarter 2020.
\bibitem{b7} P. P. Khargonekar et al., “Climate Change Mitigation, Adaptation, and Resilience: Challenges and Opportunities for the Control Systems Community," IEEE Control Systems Magazine, vol. 44, no. 3, pp. 33-51, June 2024.
\bibitem{b8} UNIFI Consortium, “UNIFI Specifications for Grid-forming Inverter-based Resources—
Version 3,” UNIFI-2026-2-1, Jan. 2026.
\bibitem{b9} G-PST Consortium, “Inaugural Research Agenda," March 2021.
\bibitem{b10} F. Dörfler, S. Bolognani, J. W. Simpson-Porco and S. Grammatico, “Distributed Control and Optimization for Autonomous Power Grids," 2019 18th European Control Conference (ECC), Naples, Italy, pp. 2436-2453, 2019.
\bibitem{b11} F. Nawaz, E. Pashajavid, Y. Fan and M. Batool, “A Comprehensive Review of the State-of-the-Art of Secondary Control Strategies for Microgrids," IEEE Access, vol. 11, pp. 102444-102459, Sept. 2023.
\bibitem{b12} J. W. Simpson-Porco, Q. Shafiee, F. Dörfler, J. C. Vasquez, J. M. Guerrero and F. Bullo, “Secondary Frequency and Voltage Control of Islanded Microgrids via Distributed Averaging," IEEE Transactions on Industrial Electronics, vol. 62, no. 11, pp. 7025-7038, Nov. 2015.
\bibitem{b13} R. Zamora and A. K. Srivastava, “Multi-Layer Architecture for Voltage and Frequency Control in Networked Microgrids," IEEE Transactions on Smart Grid, vol. 9, no. 3, pp. 2076-2085, May 2018.
\bibitem{b14} Y. Wang, T. L. Nguyen, Y. Xu, Q. T. Tran and R. Caire, “Peer-to-Peer Control for Networked Microgrids: Multi-Layer and Multi-Agent Architecture Design," IEEE Transactions on Smart Grid, vol. 11, no. 6, pp. 4688-4699, Nov. 2020.
\bibitem{b15} H. Bevrani, “Robust Power System Frequency Control," Springer, Switzerland, 2014.
\bibitem{b16} T. Huang, D. Wu and M. Ilić, “Cyber-Resilient Automatic Generation Control for Systems of AC Microgrids," IEEE Transactions on Smart Grid, vol. 15, no. 1, pp. 886-898, Jan. 2024.
\bibitem{b17} C. Kammer and A. Karimi, “Robust distributed averaging frequency control of inverter-based Microgrids," 2016 IEEE 55th Conference on Decision and Control (CDC), Las Vegas, NV, USA, pp. 4973-4978, 2016.
\bibitem{b18} C. Kammer and A. Karimi, “Decentralized and Distributed Transient Control for Microgrids," IEEE Transactions on Control Systems Technology, vol. 27, no. 1, pp. 311-322, Jan. 2019.
\bibitem{b19} S. S. Madani, C. Kammer and A. Karimi, “Data-Driven Distributed Combined Primary and Secondary Control in Microgrids," IEEE Transactions on Control Systems Technology, vol. 29, no. 3, pp. 1340-1347, May 2021.
\bibitem{b20} {\color{black}M. Maaruf, S. El-Ferik, M. Abido and M. Khalid, "Two-Loop Distributed Optimal Secondary Voltage Control of an Islanded Microgrid with Multiple Distributed Generators," 2022 IEEE PES 14th Asia-Pacific Power and Energy Engineering Conference (APPEEC), Melbourne, Australia, pp. 1-5, Nov. 2022.}
\bibitem{b21} {\color{black}M. Maaruf, S. El-Ferik, F. S. Al-Ismail and M. Khalid, "A Decentralized Multiagent-Based Robust Backstepping Control for Restoring Secondary Voltage and Frequency of Autonomous Microgrids," 2022 11th International Conference on Renewable Energy Research and Application (ICRERA), Istanbul, Turkey, pp. 430-435, Sep. 2022.}
\bibitem{b22} F. Strehle, P. Nahata, A. J. Malan, S. Hohmann and G. Ferrari-Trecate, “A Unified Passivity-Based Framework for Control of Modular Islanded AC Microgrids," IEEE Transactions on Control Systems Technology, vol. 30, no. 5, pp. 1960-1976, Sept. 2022.
\bibitem{b23} G. Cao, G. Lou, W. Gu and L. Sheng, “$H_\infty$  Robustness for Distributed Control in Autonomous Microgrids Considering Cyber Disturbances," CSEE Journal of Power and Energy Systems, vol. 10, no. 2, pp. 696-706, March 2024.
\bibitem{b24} D. D. Šiljak, “Decentralized Control of Complex Systems," Dover Publications, Mineola, NY, USA, 2012, Ch. 2.
\bibitem{b25} D. D. Šiljak, “Connective Stability of Complex Ecosystems," Nature, vol. 249, pp. 280, May 1974.
\bibitem{b26} D. D. Šiljak, “Large-Scale Dynamic Systems," Dover Publications, Mineola, NY, USA, 2007, Ch. 7.
\bibitem{b27} M. Ćalović, “Automatic generation control: Decentralized area-wise optimal solution," Electric Power Systems Research, vol. 7, no. 2, pp. 115-139, 1984.
\bibitem{b28} L. D. Marinovici, J. Lian, K. Kalsi, P. Du and M. Elizondo, “Distributed Hierarchical Control Architecture for Transient Dynamics Improvement in Power Systems," IEEE Transactions on Power Systems, vol. 28, no. 3, pp. 3065-3074, Aug. 2013.
\bibitem{b29} M. Khanbaghi and A. Zečević, “Stochastic Distributed Control for Arbitrarily Connected Microgrid Clusters," Energies, vol. 15, no. 14:5163, July 2022.
\bibitem{b30} M. Khanbaghi and A. Zečević, “An LMI-Based Control Strategy for Large-Scale Systems With Applications to Interconnected Microgrid Clusters," IEEE Access, vol. 10, pp. 111554-111563, Oct. 2022.
\bibitem{b31} J. Zhang, Y. Men, L. Ding and X. Lu, “Secondary Frequency and Voltage Regulation for Inverter-Based Microgrids: A Sparsity-Promoting DAPI Control Approach," IEEE Transactions on Control Systems Technology, vol. 32, no. 4, pp. 1512-1519, July 2024.
\bibitem{b32} F. Guo, C. Wen, J. Mao and Y. -D. Song, “Distributed Secondary Voltage and Frequency Restoration Control of Droop-Controlled Inverter-Based Microgrids," IEEE Transactions on Industrial Electronics, vol. 62, no. 7, pp. 4355-4364, July 2015.
\bibitem{b33} B. Wang and G. Verbič, “Stability Analysis of Low-Voltage Distribution Feeders Operated as Islanded Microgrids," IEEE Transactions on Smart Grid, vol. 12, no. 6, pp. 4681-4689, Nov. 2021.
\bibitem{b34} Y. Khayat et al., “On the Secondary Control Architectures of AC Microgrids: An Overview," IEEE Transactions on Power Electronics, vol. 35, no. 6, pp. 6482-6500, June 2020.
\bibitem{b35} A. Bidram, F. L. Lewis and A. Davoudi, “Distributed Control Systems for Small-Scale Power Networks: Using Multiagent Cooperative Control Theory," IEEE Control Systems Magazine, vol. 34, no. 6, pp. 56-77, Dec. 2014.
\bibitem{b36} J. Schiffer, T. Seel, J. Raisch and T. Sezi, “Voltage Stability and Reactive Power Sharing in Inverter-Based Microgrids With Consensus-Based Distributed Voltage Control," IEEE Transactions on Control Systems Technology, vol. 24, no. 1, pp. 96-109, Jan. 2016.
\bibitem{b37} M. Mesbahi and M. Egerstedt, “Graph Theoretic Methods in Multiagent Networks," Princeton University Press, 2010, Ch. 4.
\bibitem{b38} B. Abdolmaleki, J. W. Simpson-Porco and G. Bergna-Diaz, “Distributed Optimization for Reactive Power Sharing and Stability of Inverter-Based Resources Under Voltage Limits," IEEE Transactions on Smart Grid, vol. 15, no. 2, pp. 1289-1303, March 2024.
\bibitem{b39} J. P. LaSalle and S. Lefschetz, “Stability by Liapunov's Direct Method with Applications," Academic Press, New York, 1961.
\bibitem{b40} A. Colotti and A. Goldsztejn, “Practical stability and attractors of systems with bounded perturbations," 2022 IEEE 61st Conference on Decision and Control (CDC), Cancun, Mexico, pp. 5129-5134, 2022.
\bibitem{b41} E. D. Sontag and Yuan Wang, “New characterizations of input-to-state stability," IEEE Transactions on Automatic Control, vol. 41, no. 9, pp. 1283-1294, Sept. 1996.
\bibitem{b42} E. D. Sontag, “Input to State Stability: Basic Concepts and Results," in: P. Nistri and G. Stefani, "Nonlinear and Optimal Control Theory," Lecture Notes in Mathematics, vol 1932, Springer, Berlin, Heidelberg, 2008, Ch. 3.
\bibitem{b43} G. H. Hines, M. Arcak and A. K. Packard, “Equilibrium-independent passivity: A new definition and numerical certification," Automatica, vol. 47, issue 9, pp. 1949-1956, 2011.
\bibitem{b44} {\color{black}J. W. Simpson-Porco, "Equilibrium-Independent Dissipativity With Quadratic Supply Rates," in IEEE Transactions on Automatic Control, vol. 64, no. 4, pp. 1440-1455, April 2019.}
\bibitem{b45} B.T. Polyak, P.S. Shcherbakov and M.V. Topunov, “Invariant Ellipsoids Approach to Robust Rejection of Persistent Disturbances," IFAC Proceedings Volumes, vol. 41, issue 2, pp. 3976-3981, 2008.
\bibitem{b46} B. T. Polyak, A. V. Nazin, M. V. Topunov and S. A. Nazin, “Rejection of Bounded Disturbances via Invariant Ellipsoids Technique," Proceedings of the 45th IEEE Conference on Decision and Control, San Diego, CA, USA, pp. 1429-1434, 2006.
\bibitem{b47} D. D. Šiljak and D. M. Stipanović, “Robust stabilization of nonlinear systems: The LMI approach," Mathematical Problems in Engineering, 6, 810975, pp. 461-493, 2000.
\bibitem{b48} J. Simpson-Porco, F. Dörfler and F. Bullo, “Synchronization and power sharing for droop-controlled inverters in islanded Microgrids," Automatica, vol. 49, issue 9, pp. 2603-2611, 2013.
\bibitem{b49} H. M. Soliman and A. Al-Hinai, “Robust automatic generation control with saturated input using the ellipsoid method," Int. Trans. Electr. Energ. Syst., vol. 28:e2483, 2018.
\bibitem{b50} S. Boyd, L. El Ghaoui, E. Feron and V. Balakrishnan, “Linear Matrix Inequalities in System and Control Theory," vol. 15, Philadelphia, PA, USA, Studies in Applied Mathematics, 1994.
\bibitem{b51} A. I. Zečević and D. D. Šiljak, “Control of Complex Systems: Structural Constraints and Uncertainty," Springer, New York, NY, 2012.
\bibitem{b52} {\color{green}J. M. Guerrero, J. C. Vasquez, J. Matas, L. G. de Vicuna and M. Castilla, “Hierarchical Control of Droop-Controlled AC and DC Microgrids—A General Approach Toward Standardization," IEEE Transactions on Industrial Electronics, vol. 58, no. 1, pp. 158-172, Jan. 2011.}
\bibitem{b53} T. Krause, R. Ernst, B. Klaer, I. Hacker and M. Henze, “Cybersecurity in Power Grids: Challenges and Opportunities," Sensors, vol. 21, 6225, 2021. 
\bibitem{b54} A. D. Syrmakesis and N. D. Hatziargyriou, “Cyber resilience methods for smart grids against false data injection attacks: categorization, review and future directions," Front. Smart Grids, vol. 3, 1397380, May 2024.
\bibitem{b55} Z. Li, M. Shahidehpour, F. Aminifar, A. Alabdulwahab and Y. Al-Turki, “Networked Microgrids for Enhancing the Power System Resilience," Proceedings of the IEEE, vol. 105, no. 7, pp. 1289-1310, July 2017.
\bibitem{b56} A. Paudel, P. Mandal and G. Ravikumar, “Resilience Assessment of Cyber-Attacks on Distributed Secondary Control in Microgrid," 2024 56th North American Power Symposium (NAPS), El Paso, TX, USA, pp. 1-6, Nov. 2024.
\bibitem{b57} A. Bidram, B. Poudel, L. Damodaran, R. Fierro and J. M. Guerrero, “Resilient and Cybersecure Distributed Control of Inverter-Based Islanded Microgrids," IEEE Transactions on Industrial Informatics, vol. 16, no. 6, pp. 3881-3894, June 2020.
\end{thebibliography}
\end{document}